\documentclass[rmp,twocolumn,aps,nofootinbib,endfloats,showpacs,10pt]{revtex4}

\usepackage{graphicx}
\usepackage{amsmath}
\usepackage{amssymb}
\usepackage{amsfonts}
\usepackage{tabularx}
\usepackage{color}
\usepackage[normalem]{ulem}
\usepackage{dcolumn}
\usepackage{bm}
\usepackage{floatflt}
\usepackage{setspace}
\usepackage{epstopdf}

\usepackage{graphicx}
\usepackage{subfigure}
\usepackage{natbib}

\newcommand{\urusi}  {URu$_2$Si$_2$}

\begin{document}

\title{Colloquium: Hidden Order, Superconductivity and Magnetism
-- \\The Unsolved Case of URu$_2$Si$_2$}

\author{J. A. Mydosh}
\email{mydosh@physics.leidenuniv.nl}
\affiliation{Kamerlingh Onnes Laboratory, Leiden University, P.O. Box 9504,
NL-2300 RA Leiden, The Netherlands}

\author{P. M. Oppeneer}
\email{peter.oppeneer@physics.uu.se}
\affiliation{Department of Physics and Astronomy, Uppsala University, P.O. Box 516,
S-75120 Uppsala, Sweden}

\date{\today}

\begin{abstract}
This Colloquium reviews the 25 year quest for understanding the
continuous (2$^\mathrm{nd}$) order, mean-field-like phase
transition occurring at 17.5\,K in URu$_2$Si$_2$. Since ca.\ ten
years the term hidden order (HO) has been coined and utilized to
describe the unknown ordered state, whose origin cannot be
disclosed by conventional solid-state probes, such as x-rays,
neutrons, or muons. HO is able to support superconductivity at
lower temperatures ($T_c$ $\approx$ 1.5\,K) and when magnetism is
developed with increasing pressure both the HO and the
superconductivity are destroyed. Other ways of probing the HO are
via Rh-doping and very large magnetic fields. During the last few
years a variety of
advanced techniques have been tested
to probe the HO state and their attempts will be summarized. A
 digest of recent theoretical developments is also included.
It is the objective of this survey to shed additional light
on the HO state and its associated phases in other materials.\\

\end{abstract}

\pacs{71.27.+a, 74.70.Tx; 75.30.Mb}
\maketitle
\tableofcontents

\section{Historical Background}
\label{sec:hist}

Uranium is a most intriguing element, not only in itself but also
as a basis for forming a variety of compounds and alloys with
unconventional or puzzling physics properties
(for recent reviews, see  \textcite{Sechovsky98}, \textcite{Santini99}, \textcite{Stewart06}).
Natural or depleted uranium, i.e. containing 99.5 percent $^{238}$U
has a mild alpha radioactivity of 25 kBq/g, which allows U-based
samples to be fabricated and studied in university laboratories
with a minimum of safety precautions.
Following initial discoveries of unexpected superconductivity and
heavy-fermion behavior in uranium-based compounds as UBe$_{13}$
\cite{Ott83} and UPt$_3$ \cite{Stewart84}
it has become popular to synthesize uranium compounds and to cool
these in search for exotic ground states. Over the past 50 or so
years many conducting and insulating systems were synthesized,
analyzed and structurally characterized
\cite{Sechovsky98,Stewart01,Stewart06}. The usual classification
of the metallic samples at low temperature is superconducting
and/or magnetic or with some of the modern compounds designated as
``exotic'' \cite{Pfleiderer09}.

Why are uranium-based materials so interesting? The observed variety of unusual behaviors derive directly from the U open $5f$ shell.  Several
defining electronic structure quantities of the U $f$ electrons are all on the same energy scale: the exchange interaction, $5f$ bandwidth, the
spin-orbit interaction, and intra-atomic $f-f$ Coulomb interaction. As a consequence, i) elemental uranium displays intermediate behavior between
the transition metals and the rare-earths in their characteristic bandwidths, yet it generates the largest spin-orbit coupling.  ii) U lies
directly on the border between localized and itinerant (or overlapping) $5f$ wavefunctions. iii) The Wigner-Seitz radii $R_\mathrm{WS}$ of
comparative elements places U near the minimum between metallic and atomic $5f$ wave functions. And iv) Ionic U can adopt six different valences
when combined with other elements, usually one finds U$^{4+}$ with two $5f$ electrons or U$^{3+}$ with three $5f$ electrons. However, it is
difficult to distinguish or separate these two valences in metallic systems with strongly hybridized $f$ states which will then play a major role
in the ground state properties. So we now have available a weakly radioactive element that can be tuned into unique chemical and electronic states
thereby {producing} its exotic low temperature behavior. Unfortunately moving to the right along the actinide series involves strong radioactive
emissions making low temperature physical studies prohibitive except at specially equipped central facilities. The $R_\mathrm{WS}$ of elemental
uranium overlaps with that of hafnium and is yet far away from the nearly constant $R_\mathrm{WS}$ of the rare earths. Yet Hf is more a
superconducting basis than a magnetic one while U sits on the ``fence'' between superconductivity and magnetism \cite{Smith83}). For a
historical review of the actinides, see \textcite{Moore09}.

\begin{figure}
\includegraphics[width=0.99\linewidth]{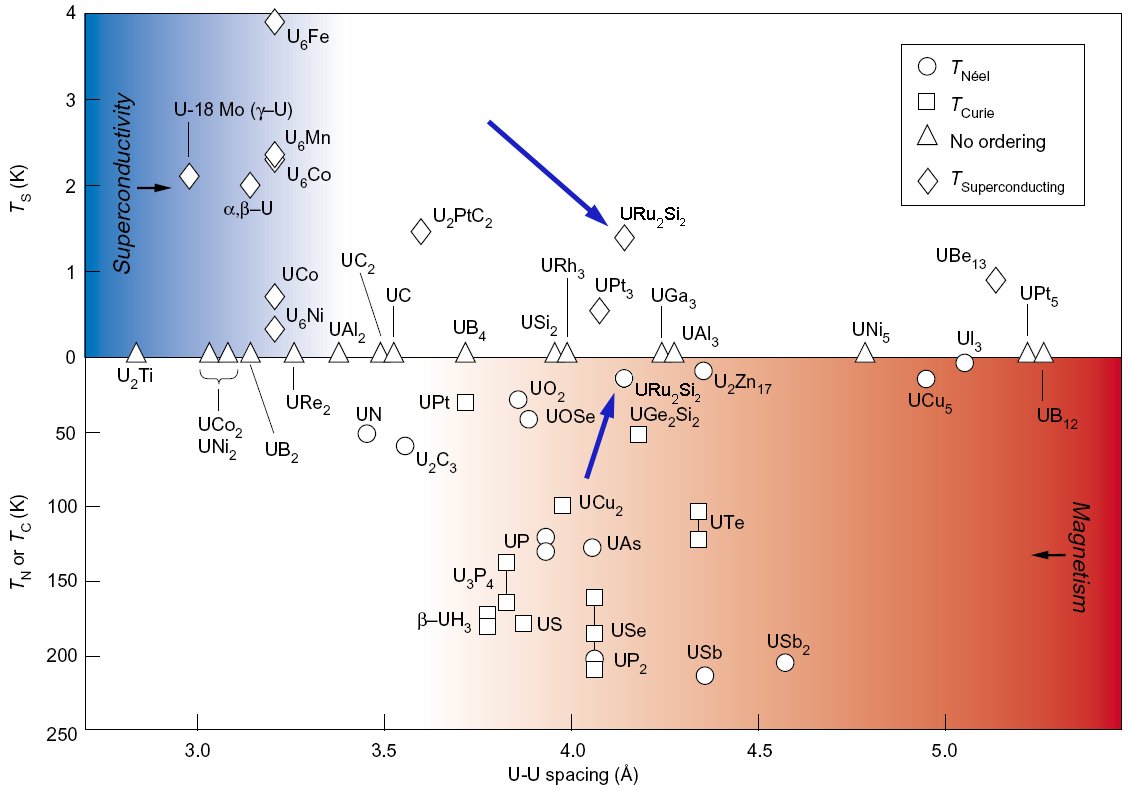}
\caption{(Color online)
The Hill plot for various uranium-based intermetallic compounds.
The bottom arrow indicates the hidden order transition temperature, $T_{\rm o}$, of URu$_2$Si$_2$
while the top one its superconducting transition temperature, $T_c$.
After \textcite{Janik08} and \textcite{Moore09}.}
\label{Hill-plot}
\end{figure}

As a traditional way of comparing the various U-based compounds
the now-famous Hill plot \cite{Hill70} is most useful. Here one
plots the ordering temperature (magnetic and/or superconducting)
against the nearest neighbor U\,--\,U spacings. According to the
trends shown in Fig.\ \ref{Hill-plot} for small U\,--\,U
distances superconducting compounds should be prominent. In the
opposite limit at large spacings greater than 3.5 \AA, significant
magnetic transitions are found. The position of URu$_2$Si$_2$ has
been added in Fig.\ \ref{Hill-plot} where the superconducting and
hidden-order transitions span the superconducting/magnetic line.
Many of the ``exotic'' or strongly correlated intermetallic
compounds, e.g., UBe$_{13}$ and UPt$_3$ do not obey Hill's rule
due to their strong hybridization of the $5f$ electrons with the
conduction electrons regardless of the U\,--\,U overlap.

Among the exotic U-based compounds there are nine unconventional
superconductors which are combined with magnetism (ferromagnetism
and antiferromagnetism) or nearly magnetic (spin fluctuations, tiny
moments or enhanced susceptibility) behaviors. For such systems the
appellation ``heavy fermion'' has been applied along with
``non-Fermi liquid'' to describe their deviant behavior. Table III in
\textcite{Pfleiderer09} surveys these materials among which UBe$_{13}$,
UPt$_3$, and URu$_2$Si$_2$ are the most perplexing.
From this table we can discern that magnetism plays a major role in
the superconductivity, sometimes generating it, sometimes
destroying it as we will see below. Here we have a comparison of
these different materials which have been the subject of
considerable research since the early 1980's \cite{Pfleiderer09}.

\section{Introduction to  ${\bf URu_2 Si_2}$}
\label{sec:intro}

\begin{figure}
\includegraphics[width=0.99\linewidth]{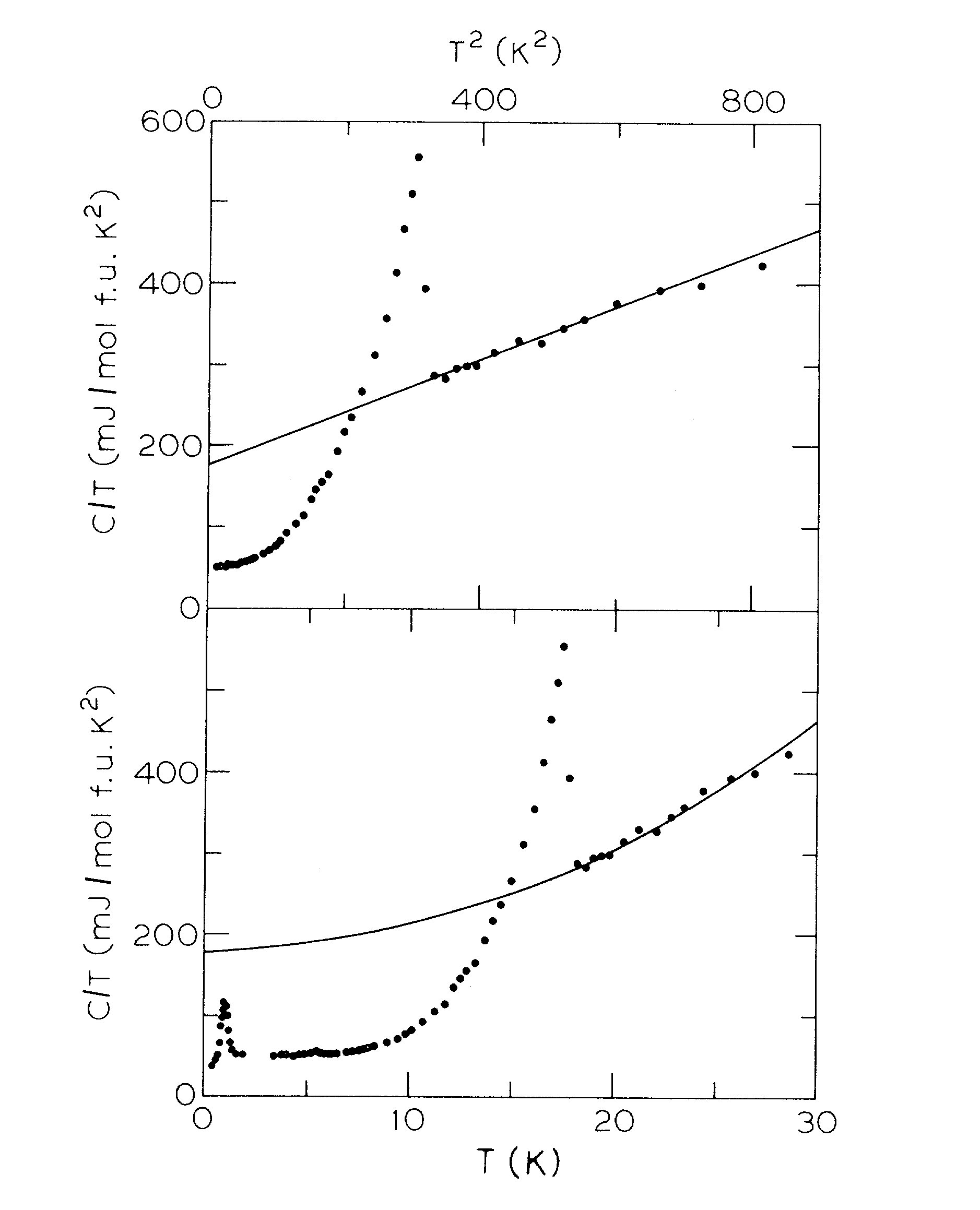}
\caption{Specific heat as function of temperature for URu$_2$Si$_2$. Top:
$C/T$ {\it vs.} $T^2$; bottom: $C/T$ {\it vs.} $T$ with the superconducting transition also shown.
Note the large extrapolated specific-heat coefficient, $\gamma$, of 180 mJ/mole\,K$^2$.
From \textcite{Palstra85}.}
\label{gamma}
\end{figure}

Only for one system, {\it viz}.\ URu$_2$Si$_2$, has there been
continuing and intense interest for the past 25 years. In 1984 a
poster by Schlabitz et al.\ (unpublished) was presented at a
fluctuating valence conference in Cologne showing the appearance
of two transitions: one superconducting, the other
antiferromagnetic. There was no publication of these results
\cite{Schlabitz86} until after two letter publications appeared in
1985 \cite{Palstra85} and 1986 \cite{Maple86}. While all three
groups agreed on the bulk superconductivity at $\approx$ 1.0\,K,
there were different interpretations for the magnetic transition
at 17.5\,K. \textcite{Palstra85} designated it a weak type of
itinerant antiferromagnetism, \textcite{Maple86} a static
charge density
wave (CDW) or spin-density wave (SDW) transition, and
\textcite{Schlabitz86} a local U-moment antiferromagnet. We now
know after 25 years that all three interpretations were incorrect.
The transition at 17.5\,K is not due to long-range ordered
magnetism and there is no measurable lattice modulation relating
to a static CDW/SDW formation. Since the origin of the transition
is unknown without a definite order parameter (OP)
established for the emerging phase or for its characteristic
elementary excitations, the term hidden order (HO) was
adapted later on for the mysterious
phase appearing at $T_{\rm o} = 17.5$\,K. Figure \ref{gamma}
illustrates the two dramatic, mean-field-like phase transitions in
the specific heat. Note the large amount of entropy forming at
$T_{\rm o}$.
 The entropy ${S} = \int_0^{T_{\rm o}} (\Delta C/T) dT$ is approximately $0.2\,R\,{\rm ln}\,2$
 ($R$ being the gas constant),
and if magnetic, this result would indicate a very large contribution
that should be detectable with magnetic neutron scattering.

\begin{figure}
\includegraphics[width=0.99\linewidth]{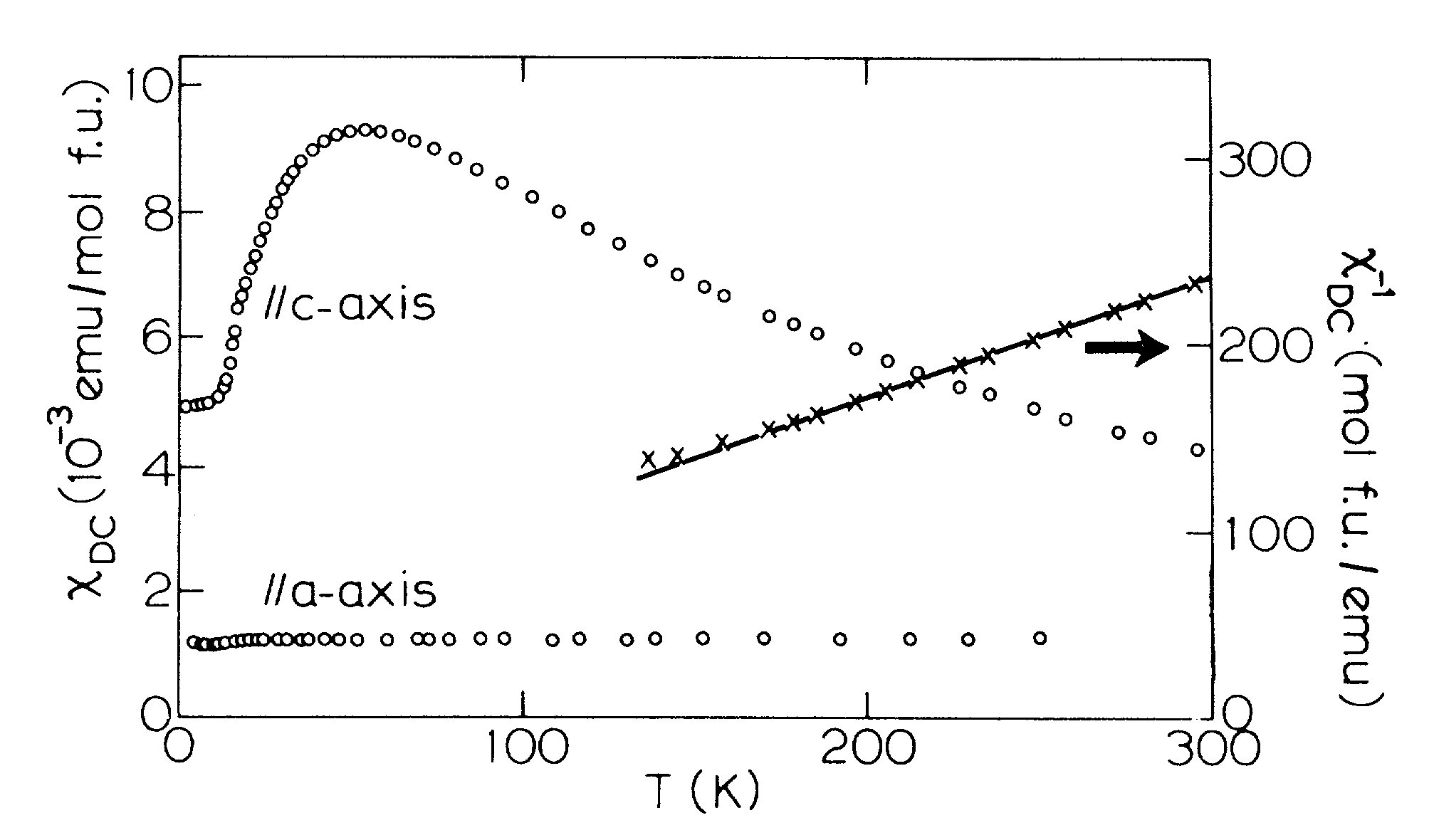}
\caption{Susceptibility {$\chi$} of URu$_2$Si$_2$ with applied field (2\,T) along the $a$ and $c$ axes.
Note the deviation from the Curie-Weiss law ($\mu_{\rm eff} = 3.5$ $\mu_B$/U; $\theta_{\rm CW} = -65$\,K)
along the $c$-axis below 150\,K. After \textcite{Palstra85}.}
\label{susceptibility}
\end{figure}

The dc-magnetic susceptibility $\chi = M/H$ with $H = 2$\,T is
displayed in Fig.\ \ref{susceptibility} for applied fields along the $a$ and $c$
axes. {The magnetic response is strongly Ising-like, as there is only
a magnetic signal}
 along $c$ which begins to
deviate from a local-moment Curie-Weiss dependence below 150\,K.
The $\chi$-maximum at $\approx$ 60\,K indicates the coherence
temperature $T^{\star}$ and the formation of a heavy Fermi liquid. The HO
transition is hardly seen but it corresponds to the intersection
of the drop with the plateau below 20\,K (cf.\
\textcite{Pfleiderer06}). Clearly the susceptibility of
URu$_2$Si$_2$ is not that of a conventional bulk antiferromagnet.
Nevertheless, there are several uranium compounds that do
show a similar Ising-like behavior, for example, URhAl and UCo$_2$Si$_2$
(see, e.g., \textcite{Sechovsky98}, \textcite{Mihalik06}).

Although the HO transition always occurs at 17.5\,K and is robust,
not dependent on sample quality, the low temperature properties
are indeed sample dependent. Especially the resistivity {$\rho(T)$} exhibits
stronger decreases as the purity of the starting U material is
increased.  A characteristic plot of $\rho(T)$ for the two $a$ and
$c$ directions in the body-centered tetragonal
(\textsc{bct}) unit cell of {\urusi} (see below) is shown in
Fig.\ \ref{resistivity}. There is the negative temperature coefficient $d\rho/dT$
at high temperatures, followed by a maximum at $\approx 75$\,K,
signaling the onset of lattice coherence, and then a
dramatic drop to low temperatures and superconductivity. Presently
one can find resitivity ratios of 500 or more with the best of
today's samples. The description of the high temperature
resistivity ($T \gtrsim 100$\,K) reaching $\approx 500$ $
\mu\Omega$-cm is open: either a strong Kondo-like scattering of
incoherent, atomic U-spins takes place or, since the resistivity
is above the Joffe-Regel limit, $k_F \,\ell \simeq 1$, (product of
Fermi momentum and  {the mean-free path}) variable range hopping
occurs. As the local U-spins disappear with the onset of
coherence when temperature is lowered and the heavy-fermion state
is created, the spin (fluctuations) scattering is removed and a
coherent low-carrier state without significant scattering is
formed. The superconducting transition temperature $T_c$ varies
significantly with sample quality and purity between $0.8$ and 1.5\,K
 and appears to coexist on a microscopic scale with the
HO without disturbing it \cite{Broholm87,Isaacs90}.

Transport and thermodynamic measurements indicated a considerable Fermi surface (FS) reconstruction occurring at the HO transition
\cite{Palstra85,Maple86}. The measured electronic specific heat in the HO and the jump in the resistivity at the transition are consistent with
the opening of an energy gap over a substantial part of the FS. \textcite{Maple86} and \textcite{Fisher90} showed that the electronic specific
heat in the HO can be extremely well described by $C_{e} (T) \propto {\rm exp}( - \Delta/ k_B T)$, where $\Delta$ is the charge gap opening in the
electronic spectrum below $T_{\rm o}$. Fits to the measured $C/T$ data gave $\Delta \approx 11$ meV and the gap was deduced to open over about
$40\%$ of the FS \cite{Palstra85,Maple86}. Subsequent resistivity \cite{Mcelfresh87} and Hall effect measurements \cite{Schoenes87} provided
additional information on the opening of a gap at the HO transition. \textcite{Mcelfresh87} deduced from resistivity measurements a gap of about 7
meV, which, with hydrostatic pressure increased to about 10 meV. The early Hall effect experiments of \textcite{Schoenes87} clearly evidenced the
opening of a gap in the HO, accompanied by a remarkable drop in the carrier concentration. These measurements also provided a rough estimate for
the single-ion Kondo temperature, $T_{\rm K} \approx 370$\,K.

Thermal expansion experiments, $\alpha$ = $L^{-1} (\Delta L/ \Delta T$), track the
{sample length $L$ and hence the lattice constants $a$ and $c$} as a function of temperature surrounding the HO transition. {The thermal expansion coefficient}  exhibits a large in-plane positive peak contrasting with the
smaller negative one along the $c$-axis, both sharply peaked at $T_{\rm o}$ \cite{deVisser86}.
Hence, there is a net volume increase indicating a
significant coupling of the HO to the lattice.

\begin{figure}
\includegraphics[width=0.99\linewidth]{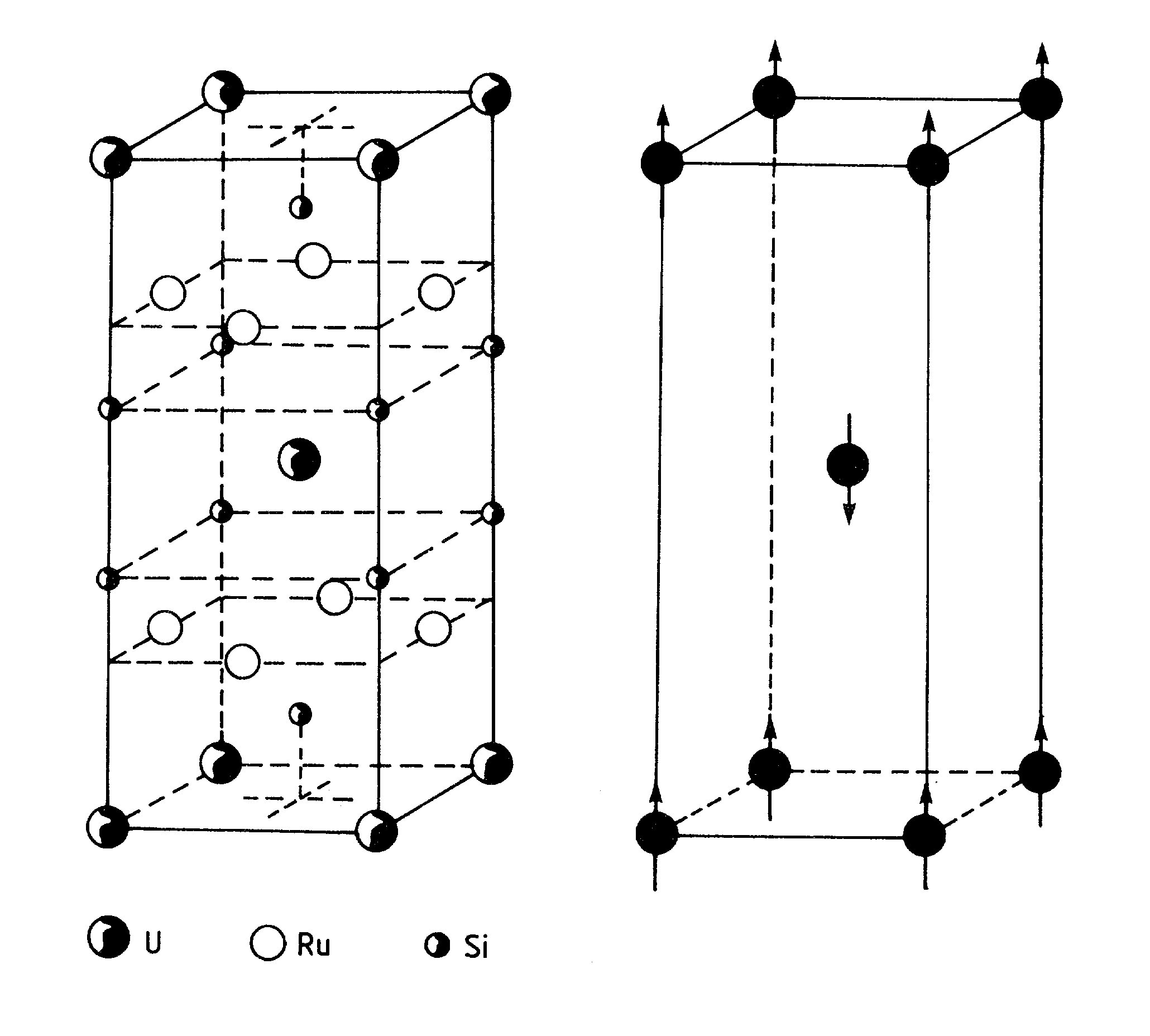}
\caption{Left: Crystal structure of body-centered-tetragonal URu$_2$Si$_2$
(space group: $I4/mmm$; lattice constants at 4.2\,K $a = 4.124$ {\AA}
and $c = 9.582$ {\AA} after a 0.1 percent
contraction from 300\,K). Right: the type-I antiferromagnetic $c$-axis spin alignment of U-moments.}
\label{structure}
\end{figure}

Because of the above conjectures of appearance of
magnetism or CDW at the HO transition, neutron scattering
\cite{Broholm87,Mason90,Broholm91} and x-ray magnetic scattering
\cite{Isaacs90} was brought to bear on the transition. By
searching for scattering in directions where Bragg peaks were
traditionally found in other compounds having the same
ThCr$_2$Si$_2$ crystal structure with known long-range magnetic
order, type-I antiferromagnetism was observed, with ferromagnetic
$a - a$ planes, alternating antiferromagnetically along the
$c$-direction
 \cite{Broholm87,Isaacs90,Broholm91}.
Figure \ref{structure} illustrates the \textsc{bct} ThCr$_{2}$Si$_{2}$ structure (space group $I4/mmm$) along with the  {putative} antiferromagnetic type-I magnetic
order on the U sublattice.
Although  magnetic Bragg peaks were found at the proper $\textbf{Q}$-values corresponding to the expected structure,
there were particular difficulties with interpreting the neutron diffraction and x-ray Bragg peaks as conventional ordered-moment
antiferromagnetism. i) The magnetic Bragg peak intensities were surprisingly small, corresponding to an ordered U moment of only $\mu_{\rm
ord} \approx (0.04 \pm 0.01)~\mu_B$/U (neutrons) and $\mu_{\rm ord} \approx (0.02 \pm 0.01)~\mu_B$/U (x-rays).
Muon spin rotation ($\mu$SR) measurements provided an ordered moment that was even an order of magnitude smaller \cite{MacLaughlin88}. ii)
The magnetic correlation lengths of about 400 {\AA} are not resolution limited, i.e., the magnetic order is not truly long-ranged. iii) The measured temperature dependence of the neutron and Bragg intensities does not resemble a typical order parameter
curve with its convex T-behavior. And iv) There are strong sample-to-sample variations of the Bragg peaks, when comparing their different
temperature dependencies below $T_{\rm o}$.
So although the conventional wisdom proposed a N\'{e}el-type magnetic
explanation even then around 1990 there were serious qualms about such
a mundane interpretation.

\begin{figure}[tbh]
   \includegraphics[bb= 0 0 490 390,clip=true,width=0.9\linewidth]{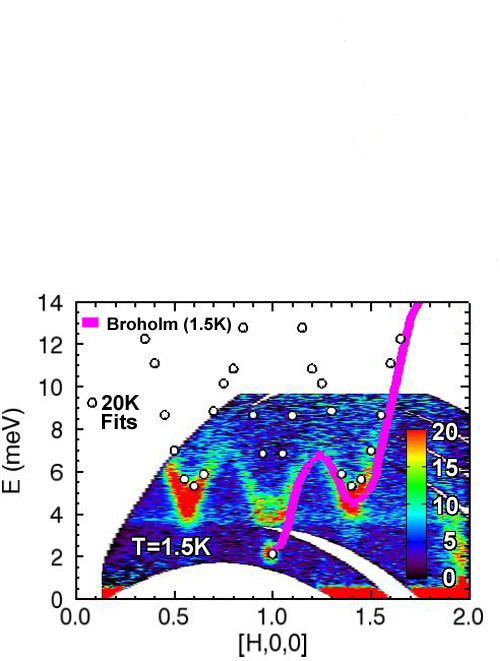}
  \caption{(Color online)
  Energy--momentum scan of spin excitations determined by inelastic neutron scattering at 1.5\,K \cite{Wiebe07}.
  The color bar denotes the intensity of the modes. The two intense modes appear at ${\bf Q}_0 = (1,\,0,\,0)$ and
  ${\bf Q}_1 = (1\pm0.4, \, 0,\,0)$ (note that the mode at $(1,\,0,\,0)$ is partially shaded by the equipment).
  The purple curve gives the magnon dispersion determined by \textcite{Broholm91}.
  Note also the commensurate gap at $\textbf{Q}_0 = (1,\,0,\,0)$ and the larger incommensutate gap
at $\textbf{Q}_1 = (1 \pm 0.4,\,0,\,0)$.
{The {\bf Q}-independent intensity detected at 3.9 meV is not due to a crystal electrical field  excitation, but due to fission processes of $^{235}$U.}
  After \textcite{Janik09}.
  \label{fig:Janik}}
\end{figure}

Early inelastic neutron work drew particular attention to magnetic excitations
appearing at low temperatures, which revealed that the inelastic neutron response of URu$_2$Si$_2$ differed
markedly from that of, e.g.,  UBe$_{13}$ and UPt$_3$
 \cite{Walter86}.
Detailed neutron experiments were performed by
\textcite{Broholm91} to detect the excitations or inelastic
modes. They observed a continuous magnetic excitation
spectrum, with two distinct gapped modes appearing at the
antiferromagnetic wavevector $\mathbf{Q}_0 = (0,\,0,\,1)$
(equivalent to $(1,\, 0,\, 0)$) and at $\mathbf{Q}_1 =
(1 \pm 0.4,\,0,\,0)$.
Figure \ref{fig:Janik} shows the  magnon energy--momentum dispersion
determined by \textcite{Broholm91} overlaid with a full inelastic
energy--momentum scan measured recently \cite{Wiebe07,Janik09}.
The modes at $(0,\,0,\,1)$ and $(1.4,\,0,\,0)$
are commensurate and incommensurate, respectively,
with the lattice. They sharply form at $T \lesssim T_{\rm o}$
with gaps of about 2 to 4.5 meV, but, since the
antiferromagnetic interpretation still prevailed in the early nineties, they
were designated as ordinary magnon modes \cite{Broholm91}.
It was noted,
though, that the antiferromagnetic mode was longitudinal and not
transversal as is commonly expected for low-energy spin
fluctuations. Thus the full meaning of these inelastic modes in
the HO was a mystery and these modes are of intense attention
today both experimentally and theoretically. Are they the
elementary excitations of the HO state that cause this spin
gapping? See Sec.\ IV below.

In order to determine the conductivity of URu$_2$Si$_2$ via a
spectroscopic probe, the far-infrared reflectance has been
measured by \textcite{Bonn88} as a function of
frequency and temperature. Using an extended Drude model they
track the optical conductivity into the coherence or heavy-fermion
regime, characterized by the reduced scattering and enhanced
effective mass, and further down below $T_{\rm o}$ where a partial charge gap opens.
This result confirms that the FS is being reconstructed
in the transition to the HO state. Hence, combining transport, thermodynamic and optical
data  with those obtained from inelastic neutron
scattering  the HO state evidently possesses both a spin and a charge
gap. Here we have the first clues of the microscopic HO behavior.

 The enigmatic nature of the HO state became first
recognized in the early nineties. The magnetic entropy $S_{m}(T)$ of {\urusi} had been determined from the $\lambda$-type anomaly observed in the
specific heat, see Fig.\ \ref{gamma} \cite{Palstra85,Maple86,Schlabitz86},  using $S_{m} (T) = \int_0^{T} (\Delta C/T') dT' $, where $\Delta C/T$
was obtained through subtracting the measured specific heat of ThRu$_2$Si$_2$, which has no $5f$ electrons, from that of {\urusi}. The entropy
formed at $T_{\rm o}$ was determined to be about 0.2\,$R$\,ln\,2 -- a relatively large value \cite{Maple86,Schlabitz86,Fisher90}. The magnetic
entropy can be expressed as $R$\,ln\,$(2S+1)$, or as $R$\,ln\,2, assuming that $N$ uranium atoms have a $S=1/2$ spin, i.e., a 1 $\mu_B$ moment. It
was noted that such a relatively large entropy change can, in particular, not be explained by assuming an antiferromagnetically ordered phase with
small moments of only 0.04 $\mu_B$ (see, e.g., \textcite{Gorkov92,Ramirez92,Walker93}). Consequently, the detected small moment antiferromagnetism
(SMAF) fails to account for the phase transition, {and} another, hidden order is responsible.

During the last two decades a quest for uncovering the HO arose. In these decades significant progress has been made in sample preparation and
characterization. Superclean {\urusi} samples were synthesized as well as samples with precisely controlled chemical substitutions. A wide variety
of experimental probes were unleashed on the HO problem. Although our understanding about the HO has definitely increased, the HO proved itself
thus far resistant. Concomitantly, many exotic theoretical models were proposed to explain the HO. Some of these could be dismissed while for
others even the experimental techniques to verify or falsify them do not yet exist. Nonetheless, the most recent experiments provided more and
clearer constraints for the theories whereby the multitude of mechanisms put forth for the HO could be narrowed down. Also, throughout these
years, 1990 to 2010, significant progress was made in characterizing the unconventional superconducting properties through a variety of
experimental probes. Experiments were carried out under pressure and with various dopings, both of which destroyed the HO and superconducting
phases. The results of these studies have been nicely summarized by \textcite{Pfleiderer09} where the recent references are included.
\\

\section{What is Hidden Order}
\label{sec:ho}

{
Hidden order evolves as a clear phase transition to a new phase at a temperature $T_{\rm o}$ as determined from bulk thermodynamic and transport measurements. Here the}
order parameter and elementary excitations of the new ordered
phase are unknown, i.e., cannot be determined from microscopic
experiments. In URu$_2$Si$_2$ all bulk quantities show a
mean-field-like (2$^\mathrm{nd}$ order) phase transition at 17.5\,K,
yet neutron and x-ray scattering, nuclear magnetic and quadrupole
resonance
(NMR, NQR), $\mu$SR, etc. are not able to reveal the OP and {elementary excitations}.
We might mention four previous cases where the OP and {elementary excitations} were only
uncovered many years after the experimental discovery:
superconductivity, antiferromagnetism, 2D X-Y magnets and spin
glass.

Basic properties of the HO state in URu$_2$Si$_2$ are i) the large
reduction of entropy upon entering it,
ii) the opening of charge and spin gaps, iii) a large decrease in the scattering rate along with a smaller effective
mass, iv) greatly reduced carrier concentration, v) a clear coupling to the lattice, and vi) an electronically ordered state that
can be destroyed by pressure, magnetic field and Rh-doping. Only out of the HO state evolves a highly unconventional ($d$-wave, even parity,
spin-singlet) multi-gap superconducting ground state with $T_c \approx 1.5$\,K \cite{Kohori96,Matsuda96}, being the focus of recent investigations
{
\cite{Kasahara07,Yano08,Okazaki08,Matsuda08,Kasahara09,Morales09}.}

Now the question arises: Is the HO phase generic, i.e., can it, as
defined above, be found in other materials with different types of
interactions? And then once defined as HO, can it be unmasked and
explained by known physical concepts and mechanisms? At present
there is no comprehensive  {understanding} of generic hidden order and its
relation to quantum criticality. Nevertheless, the concept HO is
beginning to make headway into the recent literature. For example,
in the overview article of \textcite{Coleman05} it appears to mask
the quantum phase transition in Sr$_2$Ru$_3$O$_7$ where a putative
nematic phase has replaced the HO one. Recently the term HO and
unidentified low-energy excitations were applied to the
long-standing puzzle of the low-temperature phase
transition in NpO$_2$ \cite{Santini06}. In this case  the
``unmasking'' of the HO was the identification of a
staggered alignment of magnetic multipoles. Long-range
ordering of electric or magnetic multipoles has to date only been
unambiguously detected in a few materials, NpO$_2$, UPd$_3$, and
Ce$_{1-x}$La$_x$B$_{6}$ being the most prominent examples. A
review of HO as higher-rank multipolar (non-dipolar)
driven phase transitions is given by \textcite{Santini09} and
\textcite{Kuramoto09}. A similar analogy can be drawn for the
skutterudites, e.g. PrFe$_4$P$_{12}$ and PrOs$_4$Sb$_{12}$, where
again a HO phase transition can be related to multipolar
(quadrupole) ordering and compared to URu$_2$Si$_2$
\cite{Hassinger08a,Sato08}. In addition \textcite{Dalla06} have
used HO to describe a unknown phase in a 1D Bose insulator which
exists between Mott and density wave phases; and \textcite{Xu07}
used it for unidentified phases in a quantum spin fluid.
Other areas to look for HO include the high temperature
superconductors \cite{Valla06}. { 
Here the nature of the pseudogap phase in the high-temperature cuprates
has persisted as a major unsolved problem. Recent experiments have attempted 
to expose its ``hidden order" \cite{He11}. Finally, HO remains a possibility in}
the Ce-115 compounds (hidden
magnetic order), the heavy fermions UBe$_{13}$ and UPt$_3$,
various organic charge-transfer salts near the metal insulator
transition, and non- or nearly magnetic oxides \cite{Manna10}.

The term ``hidden order parameters'' was apparently
first used in 1996 by Buyers
who was then casting doubt that the  measured small
magnetic dipole transition was truly intrinsic and that conventional
antiferromagnetism was the mediator of the phase transition at
17.5\,K. Buyers
additionally noted that inelastic neutron modes (spin excitations)
were strongly involved in the HO transition. As $T \rightarrow T_{\rm o}$
from below, the $\textbf{Q}_0 = (0,\,0,\,1)$ mode softens and its damping increases
thereby ruling out a crystal electric field (CEF) origin.

\textcite{Shah00} introduced the HO parameter in a Landau-Ginzburg free energy
expansion which tried to describe the HO phase transition via two
interacting order parameters $\Psi$, the primary unknown OP of the HO state,
 and $m$ the magnetization as determined from neutron
scattering/diffraction, as a secondary OP.  Using such
expansions \textcite{Shah00} were able to predict the magnetic
field dependence of both $\Psi$ and $m$ and thereby determine the
symmetry of the coupling between the two order parameters: $g\Psi
\cdot m$ (bilinear) or $g \Psi^2 \cdot m^2$ (biquadratic).

\begin{figure}
\includegraphics[width=0.99\linewidth]{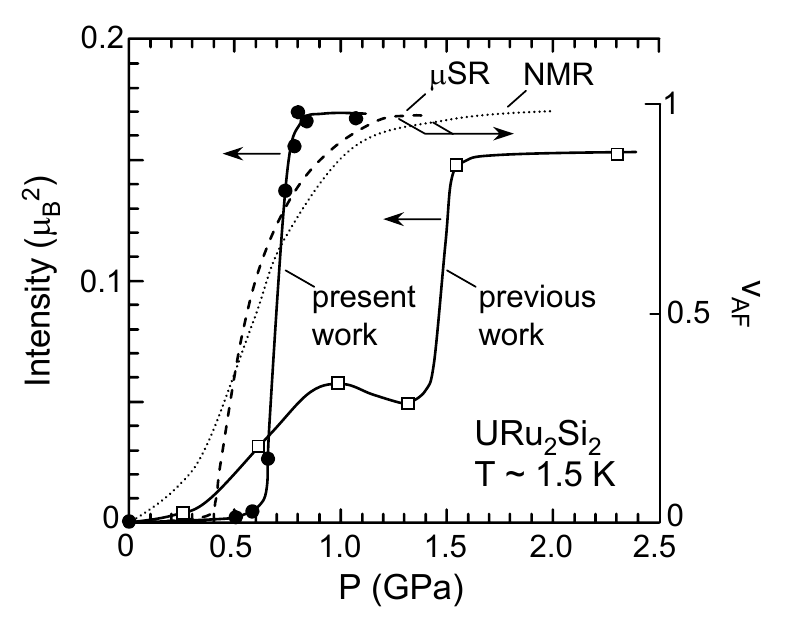}
\caption{{
Influence of the crystal purity on the phase separation between the HO and LMAF phases of {\urusi}.
Shown is the pressure dependence of the integrated intensity of the ${\bf Q}_0 = (1,\,0,\,0)$
antiferromagnetic Bragg peak measured at 1.5~K for newer high-purity single crystals (closed circles) 
and older single crystals (open squares, data from \textcite{Amitsuka99}). 
Also shown is the pressure dependence of the antiferromagnetic volume fraction (right ordinate)
derived from earlier $^{29}$Si NMR \cite{Matsuda01} and $\mu$SR \cite{Amato04} measurements. A clear phase transition
between the HO and antiferromagnetically ordered phases is  present in the purer single crystals.
After \textcite{Amitsuka07}.}}
\label{fig:Amitsuka}
\end{figure}

These two scenarios were proposed for the two types of coupling
possing the question: Does the $\Psi$ order parameter break time
reversal symmetry or not. The latter is possible, e.g. if
$\Psi$ would be
 related to a staggered electric
quadrupolar order. Based upon the measured magnetic-field dependence of the neutron moment 
\cite{Bourdarot03a,Bourdarot03b,Bourdarot05}, the comparison favored
the linear coupling scheme, thereby deducing that $\Psi$ breaks time reversal symmetry. Consequently the HO would have a limiting constraint in
its specific representation \cite{Shah00}. However, while the above deduction was sound the question still remained: Is the small magnetic moment
$m$ tracked by the neutron scattering intrinsic to the HO? This conundrum plagued the field for many years and it became a point of contention and
dispute. Only in recent years it has been fully clarified that the small antiferromagnetic moment is extrinsic, i.e., likely related to defects and stress in the sample.
{ 
Early $\mu$SR measurements provided evidence for the existence of spatially inhomogeneous HO regions and antiferromagnetic ones; the latter occupied about $10\%$ of the sample volume  \cite{Luke94}. 
Detailed $^{29}$Si NMR studies (see below) supplied definite evidence for a spatially inhomogeneous development of antiferromagnetic regions with large moments of about 0.3 $\mu_B$ \cite{Matsuda01}. 
Further arguments against a homogeneous SMAF phase were derived from dilatation experiments
\cite{Motoyama03}, $\mu$SR \cite{Amato04}, and recently, neutron Larmor diffraction  \cite{Niklowitz10}. 
The tiny staggered moment of the putative SMAF phase originates thus from the smallness of the antiferromagnetic volume fraction, a scenario which was demonstrated by the fabrication of high-purity single crystals. 
In these high-quality samples the value of the detected small antiferromagnetic moment was brought down to 0.01 $\mu_B$
and a clear phase boundary between the HO and an ordered  large moment antiferromagnetic 
(LMAF) phase was observed \cite{Amitsuka07}. Figure \ref{fig:Amitsuka} 
compares the pressure dependence of the antiferromagnetic Bragg peak at ${\bf Q}_0 = (1,\,0,\,0)$ for a high-purity single crystal (labeled ``present work") with data obtained on older single crystals (labeled ``previous work") \cite{Amitsuka07}. 
}


\section{Experimental Survey}
\label{sec:exp}

\begin{figure}
\includegraphics[width=0.99\linewidth]{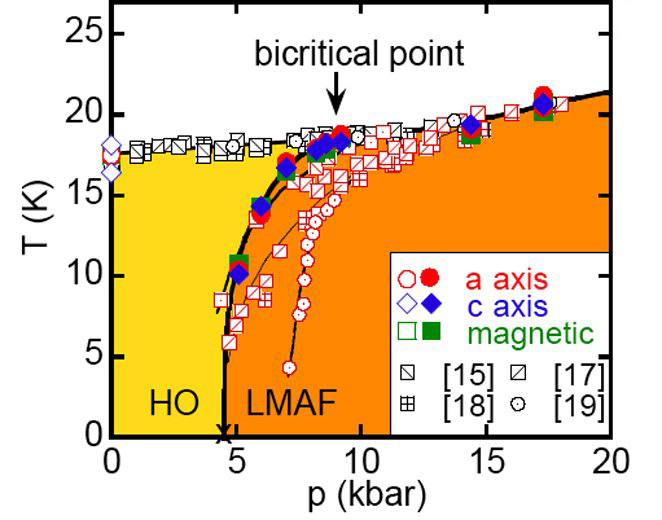}
\caption{(Color online) Collection of the {up-to-date} experimental data representing the temperature--pressure
phase diagram of URu$_2$Si$_2$ with HO and LMAF ordered states. The upper white
region is the heavy Fermi- (or Kondo-) liquid out of which the HO and LMAF develop.
[15]: \textcite{Motoyama03}, [17]: \textcite{Hassinger08b}, [18]: \textcite{Motoyama08},
and [19]: \textcite{Amitsuka07}. After \textcite{Niklowitz10}.}
\label{phase-diagram}
\end{figure}

In 1999 Amitsuka et al. investigated the pressure dependence
of the neutron magnetic Bragg peaks representing the small moment antiferromagnetism
(SMAF), with $\mu_{\rm ord} \approx 0.03$ $\mu_B$. Previous bulk resistivity measurements
at pressures up to 1.5 GPa had shown little or no change in $\rho (T)$ or $d\rho/dT$
upon entering the HO phase.  \textcite{Mcelfresh87} found  only a slight increase of
$T_{\rm o}$  as well as an increase of the transport gap.
Surprisingly the Bragg-peak intensity at ${\bf{Q}}_0 = (1,\,0,\,0)$ exhibited a dramatic
 {upturn} at a pressure of 0.5 GPa. The corresponding ordered moment was 0.4 $\mu_B$, a change
of almost 15 {(see Fig.\  \ref{fig:Amitsuka}).} This reasonably large U-moment for a heavy-fermion material
designated pressurized URu$_2$Si$_2$ to be the expected large moment
antiferromagnet (LMAF) with a conventional magnetic phase transition at
$T_{\rm N} \approx 18$\,K. Figure \ref{structure} displays the LMAF spin order. The neutron experiments
were followed by pressure dependent $^{29}$Si NMR probing the internal hyperfine fields
in both the HO and LMAF states \cite{Matsuda01,Matsuda03}. These  found evidence for a phase separation
in the HO state of a few volume percent LMAF, which increased with pressure, thereby giving the ``SMAF'' response.
It is now generally accepted that
puddles of the LMAF are generated by stress field increasing the $c/a$ axes
ratio beyond a critical value \cite{Yokoyama05}. This extreme sensitivity to sample
quality, e.g., stress, impurities, etc. has been emphasized by \textcite{Matsuda08} who
compared resistivity and specific heat data on samples cut from the middle
with those on the surface of a high-quality single crystal.
Figure \ref{phase-diagram} shows the now  {most up-to-date measurements of the}
$T - P$ phase diagram for URu$_2$Si$_2$ \cite{Amitsuka07,Niklowitz10}.
Note the first-order phase transition separating the HO from the LMAF phase.
Based upon the residual resistivity ratios a complete study of the crystal quality and various
bulk properties has been carried out by \textcite{Matsuda11}.

In order to probe the all-important Fermi surface in the HO phase of {\urusi} quantum oscillations were studied, employing both de Haas-van Alphen
(dHvA) and Shubnikov-de Haas (SdH) techniques \cite{Ohkuni99,Nakashima03,Jo07,Shishido09,Hassinger10,Altarawneh11}. The early angular dependent dHvA
measurements of \textcite{Ohkuni99} revealed three, rather small, closed FS pockets deep in the HO phase. This was once again consistent with a
substantial FS gapping occurring in the HO phase. More recently, a larger, fourth FS sheet was found by SdH measurements on an ultraclean {\urusi}
sample \cite{Shishido09}. Up-to-date angle-dependent SdH measurements revealed even a fifth branch, corresponding to a very small FS pocket {\cite{Altarawneh11}}, as
well as a previously unobserved splitting of one branch \cite{Hassinger10}, see further Sec. VII below. The quantum oscillation measurements
performed by different groups have provided data sets that are consistent with each other (cf.\ \textcite{Hassinger10}). Consequently, the FS of
{\urusi} in the HO has now definitely been established. The measured FS provides a stringent test for all theories of the HO.
A full agreement with density-functional theory (DFT) calculations has only recently been achieved \cite{Oppeneer10}. An interesting
ingredient for the HO puzzle has come from dHvA and SdH measurements under pressure, which allow to probe the FS of the HO as well as of the LMAF
phase. When pressure was applied to take the HO phase into the LMAF state, practically no change could be detected in the FS orbits as the
pressure was increased to 1.5\,GPa \cite{Nakashima03,Hassinger10}. The cyclotron effective mass decreased somewhat as expected for an increase in
the magnetic moment. Accordingly, the FS with its partial gapping is not 
{notably} modified between the HO and LMAF phases. Consistent with
this, the same FS nesting vector was detected by inelastic neutron experiments \cite{Villaume08}. Hence, the FSs of the HO and LMAF phases are
very similar, and furthermore these two distinct phases exhibit very similar transport and thermodynamical properties. This behavior has been
termed ``adiabatic continuity'' \cite{Jo07}, which however does not 
{mean} that these phases have identical OPs.


Since the quantum oscillations require a very low temperature,
usually tens of milli-Kelvin, they cannot track the entrance into
HO phase at 17.5\,K. This warrants a description of the
high temperature state out of which HO evolves. When the
temperature is reduced below 100\,K the incoherent local
moment bearing U-ions have long since disappeared due to their
hybridization with the $\textit{spd}$ electrons of the ligands.
Now a coherent heavy-fermion state is formed which in
recent nomenclature is termed the heavy-electron Kondo
liquid (KL) \cite{Yang08}. The characteristic KL temperature
(coherence $T^{\star}$) is $\approx 70$\,K in URu$_2$Si$_2$ as
determined from a variety of bulk measurements. By lowering
the temperature below $T^{\star}$ a correlation/hybridization gap
is expected to partially open at the FS due to
$f$-hybridization along this slow crossover into the KL.
In heavy-fermion materials the KL should persist down to low
temperatures (few Kelvins) where usually antiferromagnetic order
or superconductivity occurs. However, in URu$_2$Si$_2$ already at
17.5\,K the HO transition takes place with striking FS
reconstruction and gapping.

\begin{figure}
\includegraphics[width=0.99\linewidth]{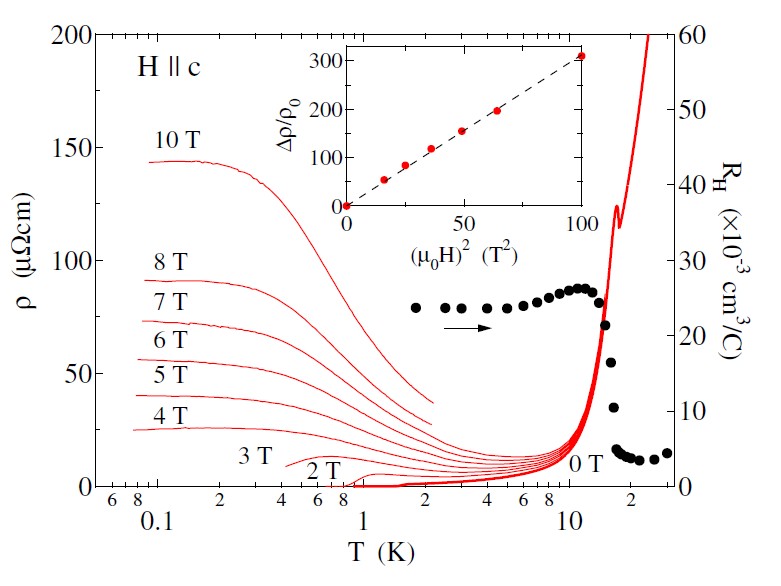}
\caption{(Color online) Resistivity as a function of log\,$T$ in various magnetic fields
(left scale). Hall coefficient through the HO transition (right scale).
Inset shows the change of resistivity versus $H^2$. From \textcite{Kasahara07}.}
\label{Hall-resistivity}
\end{figure}

Figure \ref{Hall-resistivity} collects three transport properties as the temperature is
reduced from the paramagnetic KL through the HO into the
superconducting state \cite{Kasahara07}. There is a continuous
drop in the resistivity in the HO despite the loss of carriers as
exemplified by the large jump in the Hall coefficient $R_{\rm H} \propto
1/n$, the carrier concentration. Since $\rho = m^{\star}/(e^2 n
\tau$), the scattering rate $1/\tau$ must dramatically decrease to
compensate for the reduction in $n$. Note the large residual
resitivity ratio of 670 for the  high quality URu$_2$Si$_2$
crystal {used}. The effective mass $m^{\star}$  is not expected to vary
significantly in this temperature region. The magnetoresistance
$\Delta \rho(H)/ \rho (0)$ is very large and proportional to
$H^2$, meaning that {\urusi} is a compensated electron -
hole semimetal, i.e., n$_e \approx n_h$  (as was noted
originally by \textcite{Ohkuni99}). Further analysis of
the Hall data gives a hole concentration of $\approx 0.10$ holes
per U atom in the paramagnetic KL \cite{Oh07}  {which becomes} 0.02 {holes per U atom} in
the HO \cite{Kasahara07}. It is remarkable that superconductivity
can occur at such low carrier concentration.

The thermal electricity of URu$_2$Si$_2$ is also unusual
\cite{Bel04}. There are clear indications of the HO transition
from the Seebeck coefficient (thermoelectric power) and an
unusually giant Nernst effect (ratio of transverse electric field
to longitudinal thermal gradient). These effects confirm the
drastic decrease in the scattering rate and the low density of
itinerant electrons that carry a large entropy.

Another transport property, the thermal heat conductivity $\kappa$,
displays a steep increase upon entering the HO state from the KL
\cite{Behnia05,Sharma06}. Since there are two contributions to
$\kappa$ one must separate the phonons ($\kappa_p$) from the
electrons ($\kappa_e$). As the Wiedemann-Franz law is not fully
valid here, the thermal Hall (or Righi-Leduc) conductivity
$\kappa_{xy}$ is used to independently determine $\kappa_e$. The
electronic contribution is found to be extremely small with the
phonons carrying most of the thermal conduction. This signifies
the large FS gap opening at $T_{\rm o}$ which greatly
decreases the electron--phonon scattering. By analyzing $\kappa$
in terms of the mean-field Bardeen-Cooper-Schrieffer (BCS) model,
one obtains a strong coupling of the HO with its itinerant
electrons to the lattice,  i.e., 2$\Delta (T=0) / k_B T_{\rm o} \approx 8$
 \cite{Sharma06}.
The applicability of an itinerant
model to describe the $\kappa$-data has been used to
suggest a density wave scenario with strong lattice coupling
\cite{Sharma06}. However, despite various attempts to find
periodic lattice distortions, none have been found. Solid
state probes that are sensitive to the local symmetry like NMR/NQR
would be able to detect these, but none have been found
\cite{Saitoh05}.

The elastic properties of URu$_2$Si$_2$ have been measured with ultrasonic techniques \cite{Wolf94,Kuwahara97}. A softening is observed in the $c_{11}$ and transversal $(c_{11}-c_{12})/2$ modes at temperatures below $\sim 70$ K, marking the onset of the coherence temperature \cite{Kuwahara97}. The elastic anomalies observed at $T_o$ are in contrast quite small, indicating that no uniform distortion occurs at $T_o$. The longitudinal $c_{11}$ shows a broad minimum at 30 K and increases slightly ($\sim 0.1\%$) when temperature is lowered \cite{Wolf94,Kuwahara97}.

 Point-contact spectroscopy (PCS) measurements at  temperatures around $T_{\rm o}$ have been applied to detect the opening of a conductance gap in the HO state \cite{Hasselbach92,Escudero94,Thieme95,Rodrigo97,Morales09}.
All PCS data generally evidence the opening of such gap, 
however, its onset is consistently found at about 19--25~K and not at $T_{\rm o}$. 
As explained by \textcite{Rodrigo97} the pressure that is exerted to drive the point-contact in the surface causes locally a shift of $T_{\rm o}$ to a higher value. PCS measurements at higher temperature  detected the onset of a resonance structure at the Fermi level starting at $T \approx 60$ K, consistent with the opening of a coherence/hybridization gap at $T^{\star}$ ($\approx 70$ K) derived from bulk Hall and resistivity measurements \cite{Palstra86,Schoenes87}.
\textcite{Morales09} applied PCS to the superconducting state to probe the superconducting gap below $T_c = 1.5$ K. They observe this gap, but unexpectedly and unexplained, its temperature dependence does not follow a BCS-type shape and its onset starts already above $T_c$.


Resistivity studies under pressure were performed by \textcite{Jeffries07,Jeffries08} 
and \textcite{Motoyama08}, who tracked the evolution of the charge gap
from the HO into the LMAF. The transport gap was found to increase, 
when crossing from the HO to the LMAF, with the HO gap being about 70$\%$ to
80$\%$ of the LMAF gap. Also a kink in the critical temperature was detected 
when passing from $T_{\rm o}$ to $T_{\rm N}$ under pressure.

Inelastic neutron scattering (INS) redux was carried out by \textcite{Wiebe07}
at temperatures spanning the HO transition. The authors tracked the commensurate
${\bf{Q}}_0 = (1,\,0,\,0)$ and incommensurate ${\bf{Q}}_1 = (1.4,\,0,\,0)$ modes
to higher energies and temperature. The characteristics gap energies
of 2 and 4 meV of these sharp spin waves were found in the HO phase.
Above $T_{\rm o}$ the ${\bf{Q}}_0$ mode
transforms to weak quasielastic spin flucuations. In
contrast the ${\bf{Q}}_1$ mode is due to fast,
itinerant-like, well-correlated spin excitations reaching energies
up to 10 meV, i.e., a continuum of excitations. The authors relate
these incommensurate spin fluctuations to the heavy
quasiparticles that form below the coherence temperature
$T^{\star}$ with a corresponding increase in entropy.
The gapping of such strong spin fluctuations
was related by \textcite{Wiebe07} to the loss of
entropy at the HO transition. By analyzing the spectrum for $T >
T_{\rm o}$ the specific heat could be estimated along with
the entropy loss. The calculational relationship between the spin
excitation gapping and the Fermi surface gap establishes the
strong coupling between spin and charge degrees of freedom for
this collection of itinerant electrons. Therefore, the HO
can be viewed as a rearrangement of electronic states at the Fermi energy, giving rise to
the gapping of these two coupled but distinct
types of excitations. Yet the mechanism of the HO or ``driving force'' causing the gapping
and the electronic rearrangement
remain unknown \cite{Wiebe07}. As noted here, in accordance with earlier neutron studies, no CEF excitations were found up to 10 meV.

Further INS experiments examining the intriguing inelastic modes behavior were carried
out in Grenoble \cite{Villaume08}. Here the pressure and
temperature dependence of the ${\bf{Q}}_0$ commensurate mode was determined
as the HO was entered from above. With pressure ${\bf{Q}}_0$
transforms from an inelastic excitation to a normal magnetic
Bragg peak in the LMAF, i.e., the longitudinal spin fluctuations
``freeze" and become the static long-range antiferromagnetic order, with $\mu_{\rm ord}
\approx 0.4$ $\mu_B$. In contrast the
incommensurate ${\bf{Q}}_1$ mode persists as inelastic to the highest
pressures with an increase in its energy. ${\bf{Q}}_1 = (1.4,\,0,\,0)$
represents thus a FS nesting vector found in both the HO and LMAF phases.
Conversely, the antiferromagnetic mode at ${\bf Q}_0$ is a true
signature of the HO \cite{Villaume08}. The $T -P$ phase diagram extracted
from the neutron scattering supplemented by thermal expansion and calorimetric data
\cite{Hassinger08b} is similar to that shown in Fig.\ \ref{phase-diagram}. Hence,
one can now follow the intensity of the antiferromagnetic spin
fluctuations at ${\bf{Q}}_0$ out of the KL into the HO.

\begin{figure}[tbh]
   \includegraphics[width=0.75\linewidth]{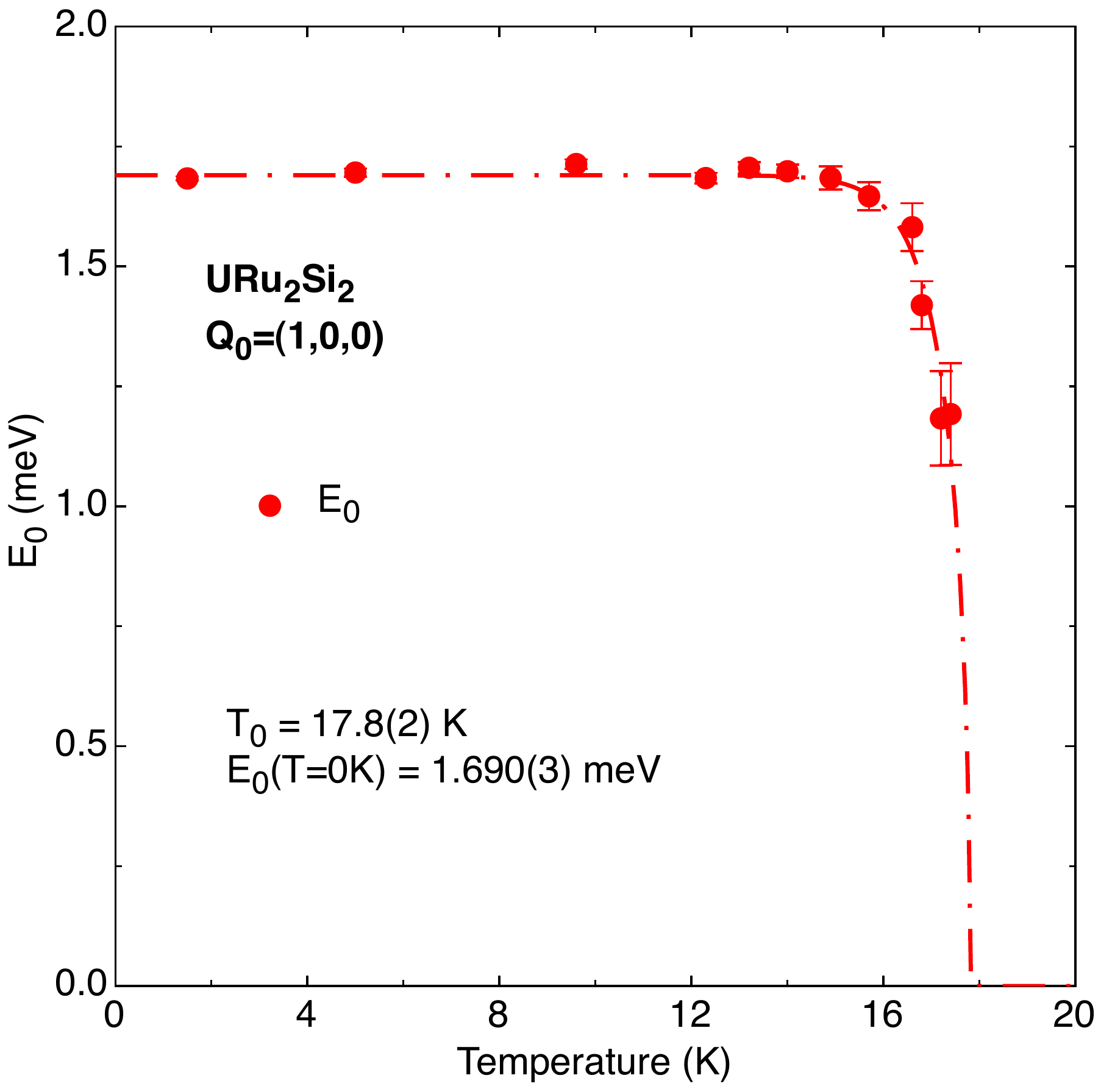}
  \includegraphics[width=0.78\linewidth]{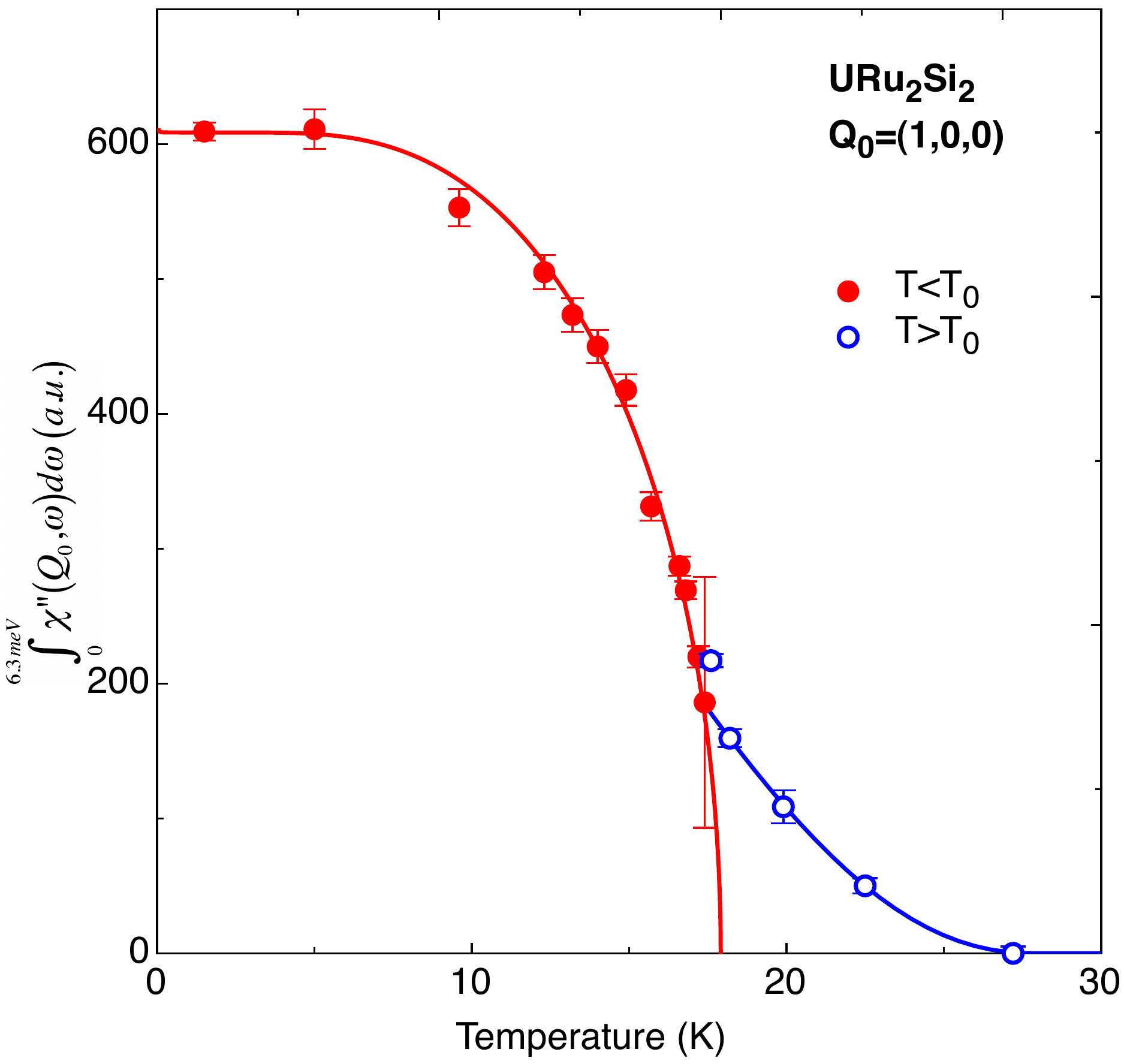}
  \caption{(Color online)  {Top}: The gapping of the spin resonance at the antiferromagnetic wavevector, ${\bf Q}_0=(1,\,0,\,0)$ {\it vs}. temperature. Note the sudden onset of the spin gap $E_0$ upon entering the HO state.
   {Bottom}: The integrated dynamical spin susceptibility of the antiferromagnetic spin resonance {\it vs.} temperature. The integrated intensity displays order parameter behavior and thus tracks the OP of the HO.
 After \textcite{Bourdarot10}.
   \label{fig:AF-OP}}
\end{figure}

\begin{figure}
\includegraphics[width=0.8\linewidth]{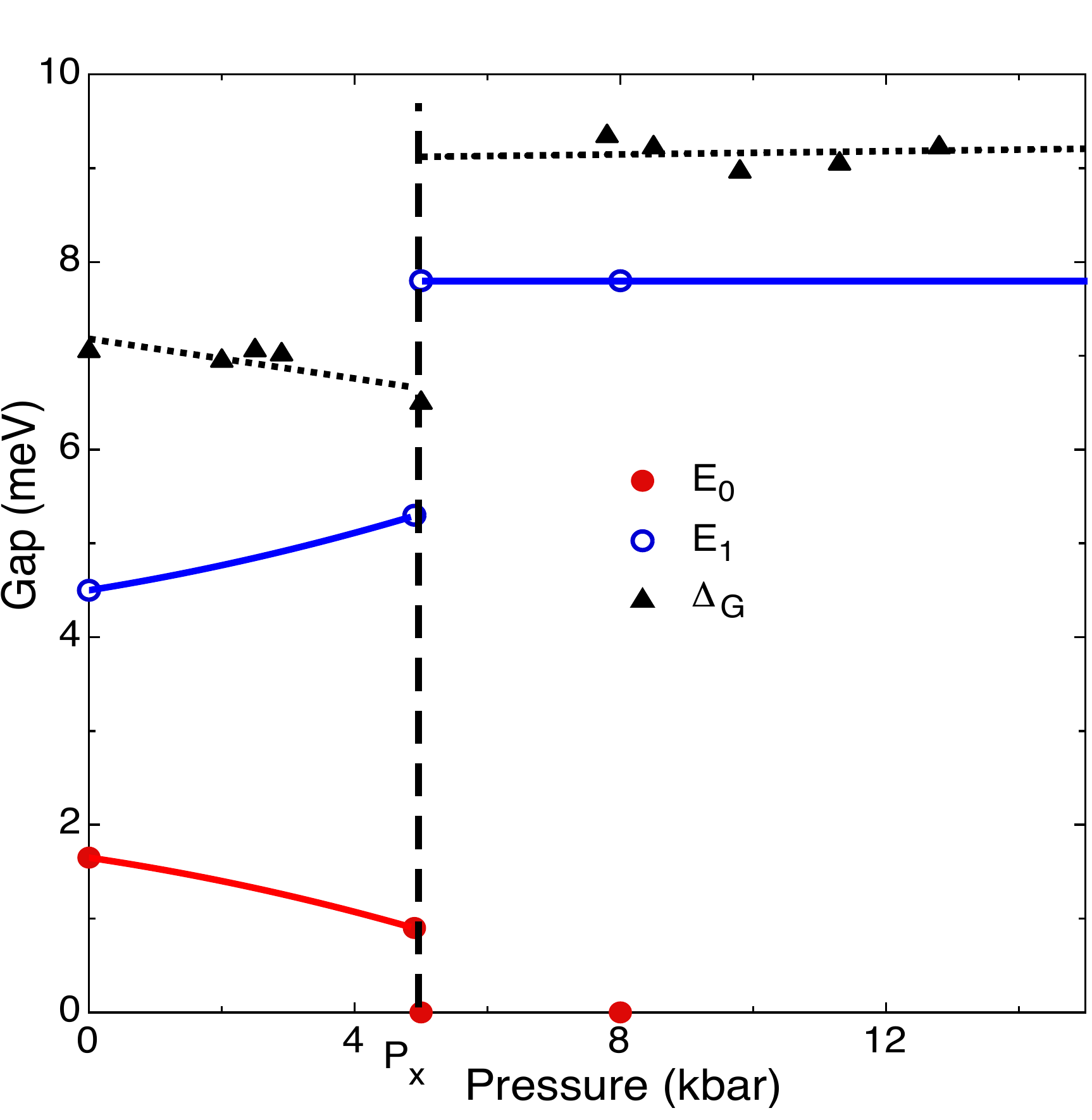}
\caption{(Color online)
Energy gap scales of HO and LMAF phases determined from INS resonances and
bulk resistivity measurements as function of pressure
spanning HO to LMAF. $\Delta_{\rm G}$ is the transport gap, whereas $E_0$ and $E_1$ are
the gaps of the commensurate and incommensurate spin resonances, respectively.
After  \textcite{Bourdarot10}.}
\label{gaps}
\end{figure}

Very recently a detailed study of the (polarized) inelastic neutron resonances at ${\bf{Q}}_0$ (and ${\bf{Q}}_1$) was performed by the Grenoble
group focusing on the commensurate antiferromagnetic resonance  at (0,\,0,\,1) \cite{Bourdarot10}. The dynamical spin susceptibility
$\chi({\bf{Q}},\omega,T)$ that is related to the spin-spin correlation function was determined and analyzed above and below $T_{\rm o}$. Clear
spin correlations at ${\bf{Q}}_0$ were observed as the temperature was scanned through $T_{\rm o}$ with a jump in the resonance energy or spin-gap
energy $E_0$ and a BCS-like T-behavior of the integrated imaginary susceptibility, see Fig.\ \ref{fig:AF-OP}. It was predicted by
\textcite{Elgazzar09} that the integrated spin-spin correlation function should display OP behavior.  It thus appears that the
${\bf{Q}}_0$ spin resonance
is a main signature of the HO phase. As
proposed by \textcite{Wiebe07} the incommensurate
${\bf{Q}}_1$ mode that appears in both the HO and LMAF phases
accounts for a large share of
 the loss of entropy due gapping of the mode and corresponding
 loss of spin fluctuations.
 A major question is how the spin gap is connected to the charge gap of the FS.
Is the spin resonance at the antiferromagnetic wavevector driven by the electronic gapping of the FS, or {\it vice versa} is the spin resonance
responsible for the FS gapping? In order to explain the observed behavior \textcite{Bourdarot10} suggest that that both itinerant and local $5f$
electrons are playing a role, with itinerant electrons responsible for the spin gaps and localized electrons for the FS gap, thereby requiring a
duality interpretation. Figure \ref{gaps} collects the various energy scales for the neutron resonances and compares them with the gap energies
from bulk resistivity measurements. Note the step-wise increase of the charge gap $\Delta_{\rm G}$ at the HO--LMAF phase transition. The
spin gap $E_1$ of the incommensurate mode also step-wise increases, whereas the commensurate mode with gap $E_0$ vanishes in the LMAF.

So with this collection of sophisticated experiment on ever improving crystal quality,
what have we learned about HO? We will now give a brief summary to conclude this section.

HO is not a magnetic dipole (local-moment) ordered transition and
ground state. Yet as with all strongly correlated electron systems
magnetism hovers in the background, every ready to make an
appearance, in our case as dynamical spin excitations. The salient
features of the HO are the pronounced Fermi surface reconstruction
and gapping at $T_{\rm o}$. Surprisingly, there is practically no
change in the FS properties between HO and LMAF.
The high-T phase
out of which HO appears can be designated as the recently proposed
Kondo liquid \cite{Yang08}. This {raises} the question as to the exact
nature of a Kondo effect in URu$_2$Si$_2$. Above $T_{\rm o}$ a
correlation/hybridization gap slowly opens in the coherence
crossover regime, its importance has up till now been neglected.
The FS topology of {\urusi} in the HO has now been
determined to consist of only  relatively small, closed pockets.
The HO ground state is that of a very low carrier concentration,
compensated semimetal with strong lattice coupling. Itinerant
electrons appear to be the main characters here as most experimental
evidence supports an itinerant electron or band-like scenario.
There is no ``smoking gun'' for a localized uranium $5f^2$ or
$5f^3$ configuration (see \textcite{Oppeneer10} for a
discussion of this issue).

The INS has given us two modes or resonances: Preeminent is the
${\bf{Q}}_0$ commensurate resonance which mimics OP behavior of
the HO state. The ${\bf{Q}}_1$ incommensurate spin mode, present
in both the HO and LMAF phases, could be responsible for the
similar loss of entropy at $T_{\rm o}$ in HO and at $T_{\rm N}$ in LMAF
\cite{Wiebe07,Balatsky09}. There is an intimate relationship
between charge and spin degrees of freedom. A special charge-spin
duality or coupling seems to exist here analogous to what {is} now
fashionable in the topological insulators,
{where a similar charge-spin duality comes into play \cite{Hasan10}.}
Nevertheless, the
driving force or mediator for creating the HO transition remains
to be clarified.


\section{Theoretical Survey}
\label{sec:theo}



There have been a large number of theoretical contributions to the
HO problem spanning the past 25 years. We review these here and
attempt to relate them to the present experimental developments.

The earliest theoretical models focussed on explaining the
unusually small ordered moment and its Ising-like behavior that
had been found in the neutron scattering and susceptibility
experiments \cite{Broholm87,Palstra85}. A first  CEF model for
{\urusi}  was developed by \textcite{Nieuwenhuys87}, who
considered a U$^{4+}$ -$5f^2$ ion with total angular momentum
$J=4$ in a tetragonal crystal field. He showed that a reasonable
agreement with experimental susceptibility data could be obtained
when the three lowest CEF levels would be singlets with a
splitting of about 40 K between the ground and first excited state
($|\Gamma_1^{(1)}\rangle =\alpha (|4\rangle +|-4\rangle) +\beta
|0\rangle$ with $2\alpha^2 +\beta^2=1$ and $|\Gamma_2 \rangle =
(|4\rangle -|-4\rangle)/\sqrt{2}$, respectively). The predicted
ordered moment was however ten times larger than the measured one.

Two early on theoretical models began to treat the HO phase beyond the simple small-moment antiferromagnetic ordering. To explain non-N\'{e}el
magnetically ordered phases characterized by weak antiferromagnetism \textcite{Gorkov91} and \textcite{Gorkov92} proposed double ($\langle
S_{\alpha} (\boldmath{r}_1) S_{\beta} (\boldmath{r}_2) \rangle$) and triple
 ($\langle S_{\alpha} (\boldmath{r}_1) S_{\beta} (\boldmath{r}_2)  S_{\gamma} (\boldmath{r}_3)\rangle$)
spin correlators as driving order parameter which break spin rotational symmetry,
yet the local magnetization is zero. Here the spins can be on different U ions
or on the same U ion, in which case the spin correlators are equivalent to quadrupole or octupole moments.
\textcite{Ramirez92} analyzed the nonlinear susceptibility and proposed a double-spin correlator,
that would give rise to a staggered quadrupolar order.

\textcite{Walker93} performed angular dependent {\it polarized}
neutron scattering to search for higher order spin correlators and
spin nematic order parameters. A substantial neutron spin-flip
scattering was observed, which was interpreted as evidence against
a spin nematic or quadrupolar magnetization distribution. Various
higher order multipoles could be excluded as well. It is however a
question in how far these measurements were influenced by a
parasitic SMAF phase. As a test of the proposed non-N{\'e}el
orders \textcite{Barzykin93} predicted that there should exist
broken-symmetry Bragg peaks for spin nematic and triple-spin
correlator phases that could be induced with an external magnetic
field.  Neutron experiments searching for such a field dependence
were undertaken but gave a null result, thereby discarding these
theories \cite{Mason95,Buyers96}.

An extensive CEF treatment was developed by \textcite{Santini94}, who considered the same U$^{4+}$ CEF Hamiltonian as \textcite{Nieuwenhuys87},
but analyzed in detail the possible energetic orderings of the nine CEF states and considered different angular momentum operators that can
support multipolar order parameters. They observed that there exist three possible variants for the three lowest singlet levels. The first variant
(A), that was considered by \textcite{Nieuwenhuys87}, could support dipole order ($J_z$)  {simultaneously} with hexadecapole  order ($J_xJ_y[J_x^2 -
J_y^2]$), but this variant was rejected because it would give a much too high Schottky contribution to the specific heat. The other two variants
(B and C) could both sustain a quadrupole ($J_xJ_y$ or $(J_x^2-J_y^2)$) and an octupole ($J_z(J_x^2-J_y^2)$ or $J_zJ_xJ_y$) order parameter.
Based on a comparison with experimental data \textcite{Santini94}
favored electric quadrupolar ordering of localized $f$ electrons for the HO.
The dipole matrix elements would be  zero for these two variants, but they suggested that
a small static dipole magnetic moment could be induced by quadrupolar ordering.
\textcite{Walker95} however noted that dipole order cannot be induced as a secondary
OP by quadrupolar order and that the proposal would not be compatible
with the polarized neutron scattering (see also \textcite{Santini95,Santini98}).
As mentioned before, there is to date no experimental evidence of such localized CEF levels and splittings.

A different CEF interpretation was proposed by \textcite{Barzykin95}
who suggested that U ion could be in an {\it intermediate valence}
configuration, dominantly U$^{4+}$, but with a small admixture of
half-integer spin configurations ($5f^1$ or $5f^3$) being responsible for the small moment.

The first suggestion of an itinerant-localized duality model was
made by \textcite{Sikkema96}, who developed an Ising-Kondo lattice
model, in which they adopting the U$^{4+}$ two singlet CEF level ground state
(variant A) with additional on-site exchange coupling to the
conduction electron spins, leading to Kondo screening. In this
model the HO is explained as small moment antiferromagnetism, but
the appearance of such a state is now regarded as being parasitic
to the HO.
Another itinerant-localized duality model was
suggested by \textcite{Okuno98}. They used an induced-moment mechanism
for the SMAF appearing from a singlet-singlet CEF scheme along
with a partially nested FS of the itinerant electrons.
Accordingly, the SMAF is found to be compatible with the large
specific heat jump and is composed of both spin and orbital
components. While the duality model is also employed in more
current HO theories, see for example \textcite{Tripathi05}, here
it is focused on explaining the SMAF which is now believed
to be extrinsic.

\textcite{Kasuya97} proposed that the HO would be due to
dimerization on the U sublattice. Here the corresponding atomic
displacements should be observable with x-ray diffraction or
extended x-ray absorption fine structure (EXAFS) but have never
been detected.


Another novel starting point for the HO transition is the
unconventional spin density wave (SDW) theory by
\textcite{Ikeda98}. In this model the electron-hole pair amplitude
changes its sign in momentum space. For the $d$-wave
superconductors this is called a $d$-density wave (DDW). Using a
mean-field extended Hubbard model, the energy gap and
thermodynamic properties can be calculated. They exhibit a sharp
specific-heat cusp, a clear drop in the uniform susceptibility and
a zero staggered magnetization. Except for the susceptibility the
above temperature dependence agree with experiment. However, a
microscopic signature of the DDW has not yet been found.
Presumably an ARPES measurement would be able to detect the FS
modulation in $\textit{k}$ space. A modern extension of the DDW
model is the chiral DDW of \textcite{Kotetes10a} and
\textcite{Kotetes10b} who demonstrate that a chiral ground state
could be responsible for the anomalous thermoelectric and
thermomagnetic effects (e.g., Nernst effect) in URu$_2$Si$_2$.

As mentioned above, different CEF schemes can be adopted
for the U$^{4+}$ ion. Experiments performed on diluted
U$_x$Th$_{1-x}$Ru$_2$Si$_2$ with $x \le 0.1$ (see below) were
interpreted as an indication of a \textit{doublet} CEF ground
state \cite{Amitsuka94}. If the  $5f^2$ CEF ground state
is assumed to be a non-Kramers  $\Gamma_5$ doublet
($|\Gamma_{5\pm}\rangle = \gamma |\pm3\rangle +\delta
|\mp1\rangle$, $\gamma^2 +\delta^2=1$) in stoichiometric {\urusi}
as well, different  multipole characters could become
possible. \textcite{Ohkawa99} considered the $\Gamma_5$ doublet
and from a consideration of angular momentum matrix elements
proposed that a magnetic dipole or a non-magnetic quadrupole can
be formed. Depending on the fine tuning (e.g., pressure) one can
transform one ground state into the other. So within this local
CEF scheme the HO and the LMAF phases are explained as
quadrupolar {and} spin-dipolar ordering, {respectively}. 
Once again, the
local CEF levels have never been experimentally observed -- they
should be hybridized away in a KL. Yet,  based upon an analysis of
the dHvA amplitude \cite{Ohkuni99} as a function of magnetic field
orientation a series of 16 nodes appear. This behavior can be
related to an Ising degree of freedom in the U$^{4+}$ electronic
configuration, that would be possible with a $\Gamma_5$
doublet. According to \textcite{Silhanek06b} the local $\Gamma_5$
doublet becomes hybridized with the conduction electrons and
itinerant quasiparticle bands having $\Gamma_5$ character are
formed via this hybridization. This model then proposes
itinerant antiferro quadrupolar order for the HO.
Although such electrical quadrupoles could be detectable
with resonant x-ray scattering (RXS) they have thus far not been observed
{\cite{Amitsuka10,Walker11}}.

 \textcite{Yamagami00} proposed that the HO could be an
antiferromagnetic phase with a small magnetic moment caused by
cancellation of a large spin and an equally large but antiparallel
orbital moment. Neutron form factor measurements would be able to
detect this, but could not confirm this effect (see, e.g.,
\textcite{Broholm91,Kuwahara06}).

Within the basic ingredients of a heavy Fermi liquid,
\textcite{Chandra02} proposed that the HO in URu$_2$Si$_2$ resulted
from {\it incommensurate} orbital antiferromagnetism associated
with circulating charge currents between the U-atoms. The model
was based upon the appearance of a weak internal field below
$T_{\rm o}$ detected by the $^{29}$Si NMR line-width broadening
\cite{Bernal01}. These plaquette currents cause an orbital moment
to form, thereby breaking time reversal symmetry, that can be
estimated from the isotropic field distribution at the Si sites.
Thus the HO order parameter is taken to be proportional to the
temperature onset of the line-width broadening.
\textcite{Chandra02} calculated the local fields and the large
entropy transition in the specific heat, and detailed predictions
were made to relate the incommensurate current ordering to the
neutron-scattering cross-section. A ring of scattering intensity
was forecast at a specific anisotropic $\bf{Q}$-radius which was
experimentally sought but not found \cite{Wiebe04}. In addition
the NMR line-width broadening was found to be much reduced
when higher quality stress-free single crystals were investigated
\cite{Bernal06,Takagi07}.

Scenarios for higher-order multipolar transitions were put
forward by \textcite{Kiss05} and \textcite{Fazekas05}. By
assuming an itinerant to localized transition upon entering the HO
phase from above, \textcite{Kiss05} treated the HO state as
multipolar orders via symmetry-group theoretical arguments. Within
the 3-singlet CEF model, U$^{4+}$ - $5f^2$ ions can carry a
sequence of higher multipole order parameters. The authors thereby
consider a manifold of multipolar order parameters spanning the
gamut from magnetic dipoles (rank 1) to magnetic triakontadipoles
(rank 5). Led by the time-invariance breaking detected by
neutrons \cite{Walker93,Bourdarot03a}, by NMR \cite{Bernal01,Bernal06},
and $\mu$SR \cite{Amato04} they propose staggered octupolar order
for the HO, which would be time-reversal symmetry breaking and
have a vanishing total dipolar moment.  Further arguments
were based on comparisons with uniaxial stress measurements, from
which they concluded that a triakontadipole (which also breaks
time-reversal symmetry) would be incompatible with the uniaxial
response to stress.
Notably, \textcite{Fazekas05} predict the onset of octupolar order at $T_{\rm o}$, followed by a second transition due to quadrupolar ordering at $T \approx 13.5$\,K.
While this theoretical conjecture is most intriguing, the observation of multipoles of rank higher than two
is experimentally hardly accessible. Also, there is little evidence for another
transition below $T_{\rm o}$. Nevertheless this work has inspired a succession of
additional theoretical efforts directed toward higher-order multipoles.
\textcite{Hanzawa05} adopted a singlet-doublet CEF level scheme and deduced that $J_x (J_y^2 -J_z^2)$ octupolar order would be most probable.  {\it Incommensurate} ordering of octupoles
was suggested by \textcite{Hanzawa07}.
Very recently, an antiferro ordering of time-even hexadecapoles was proposed as HO parameter by \textcite{Haule09} and a ferromagnetically aligned {(${\bf Q}=0$)} time-odd triakontadipolar HO parameter was proposed by
\textcite{Cricchio09}. Antiferro-quadrupole order of $J_xJ_y$-type was proposed by \textcite{Harima10}, and antiferro-hexadecapolar ordering of $xy(x^2-y^2)$ type was proposed by \textcite{Kusunose11}.
A full review of multipole orders in strongly
correlated electron systems has recently been given by \textcite{Santini09} and \textcite{Kuramoto09}.

A serious drawback of the CEF multipolar theories is that
these from the outset assume a localized $5f^2$ configuration, but
as mentioned before, there is no experimental evidence for such
localized CEF levels and splittings. Moreover, recent resonant
x-ray scattering experiments sought for quadrupolar order
but could not observe it \cite{Amitsuka10,Walker11}. This would thus exclude any quadrupole scenario. Also as discussed by \textcite{Walker11} the magnetic formfactor of {\urusi} is normal,
which would exclude magnetic octupolar and triakontadipolar orders. Conversely,
previous RXS measurements on established multipolar
ordered materials did, for example, unambiguously detect
 non-collinearly staggered quadrupole
order in NpO$_2$, Ce$_{1-x}$La$_x$B$_6$, and UPd$_3$
(see \textcite{Paixao02}, \textcite{Walker06}, and discussions provided by
\textcite{Santini09,Kuramoto09}). Also, NMR measurements have been
shown to be able to detect quadrupolar symmetry breaking in NpO$_2$
(see, e.g., \textcite{Tokunaga05}), although it could be that
contributions cancel out in {\urusi}. An often neglected aspect of
CEF theories is that they do offer insight in the possible
multipole  symmetries on a single U ion, but this is not
sufficient to achieve collective long-range order, which requires
a mechanism providing exchange coupling of multipoles.
Investigations of the multipolar exchange interaction are only in
their initial phase (cf.\ \textcite{Suzuki10}). Lastly, while thus far all
CEF theories assume a U$^{4+}$ ion, a recent electron energy loss
spectroscopy (EELS) study determined a $5f$ occupation of 2.7
\cite{Jeffries10}, which is not close to a $5f^2$ configuration.
{The EELS result, combined with a magnetic 
entropy approaching $S_{mag} \sim R \,{\rm ln} 4$ at high temperatures \cite{Janik08},
would rather suggest a Kramers $5f^3$ ion to be realized at high enough temperatures. 
As mentioned before at low temperatures itinerant $f$ character appears to prevail. }

Helicity order, i.e., the establishment of a fixed axis of
quantization for the spins on the FS, was proposed as the HO cause
by \textcite{Varma06}. Such order arises when the Pomeranchuk
criteria for the spin-antisymmetric Landau parameters are violated
with respect to the Fermi-liquid state. In order to remove or
``cure'' this instability a nematic phase transition occurs
thereby creating the HO phase. For URu$_2$Si$_2$ this model
represents a displacement of the FS into up and down sheets. Based
upon the then-known band structure and FS, and variations of the
density of states near the FS, \textcite{Varma06} calculated
different experimental features: specific heat, linear and
non-linear susceptibilities, and the NMR line-width broadening.
Using the properly modified Landau parameters good agreement with
the data was found. The main difficulty with this approach is the
neglect of the exact, complex FS and its significant hot spot 
gapping upon
entering the HO state. Also, as spin-orbit coupling is
strong in uranium, a picture of pure spin up and down states
is not sufficient. Positron annihilation was employed to
study FS changes in the HO, but could not confirm helicity order
if this state would form as a single domain \cite{Biasini09}. Up
until now the suggested microscopic experimental verifications 
\cite{Varma06} of
the helicity order have not been accomplished.

By way of relating the HO transition to a SDW \textcite{Mineev05} employed a double dual approach whereby two order parameters $\Psi$ and $m$ were
considered. In addition there were also two subsystems: local CEF moments on the U$^{4+}$ sites along with conduction electrons in nested bands.
Strong electron-electron interactions were added to drive the SDW in a commensurately nested FS. The SDW then induces a local SMAF or $m$ in the
HO state. $\Psi$ represents the SDW amplitude and is the primary order parameter which governs the experimental behavior. The pressure dependence
is predicted to show a line of first-order transitions ending in a critical end-point that separates the SMAF from the LMAF. The high magnetic
field properties are calculated from the model to account for the multiple phases (See Sec.\ VI). Unfortunately experiment does not find a SDW
even with a small form factor and it is now commonly accepted that the SMAF is extrinsic, and instead of a critical end-point there is the merging
of all three phase lines at a bicritical point \cite{Motoyama03,Amato04,Amitsuka07,Hassinger08b,Niklowitz10}. Figure \ref{phase-diagram} presents
the up-to date pressure-temperature phase diagram for URu$_2$Si$_2$. Note the three distinct phases: KL, HO and LMAF. The SMAF is omitted since it
is extrinsic.

\section{High Magnetic Fields and $\bf Rh$-Doping}
\label{sec:fields}

There were various pioneering attempts to study the high magnetic-field
$\mbox{\boldmath$H$}$ behavior of URu$_2$Si$_2$ in the HO state and beyond by
\textcite{Deboer86} and \textcite{Sugiyama90}. Such experiments showed the destruction
of the HO phase transition with fields approaching 35\,T and the
occurrence of novel phases between 35 and 42\,T. However, the use of
short-time ($<$1\,ms) pulsed fields resulted in a loss of temperature equilibrium
for these highly conducting samples and the $T - H$ phase diagram
was unclear. Here the field $\mbox{\boldmath$H$}$ is applied along the tetragonal $c$-axis,
with little or no magnetic response for fields in the basal plane.

More recent investigations began in 2002 with continuous $H$-fields
reaching 45\,T for specific heat, magneto-caloric effect (MCE), and magnetoresistance
\cite{Jaime02,Kimj03}. These measurements detected the suppression
of the HO phase at 35.9\,T ($H_0$), the appearance of a new phase between 36.1\,T
($H_1$) and 39.7\,T ($H_2$)
with no additional phase transition above 40\,T. While the low field HO
transition in the specific heat is $\lambda$-like (second-order), with
increasing field its shape changes to a symmetric peak indicating a first-order
transition. The new high-field phase is always reached through a first-order
transition. The above results are confirmed via the MCE, which uncovers an
increase in the magnetic entropy at $H_0$, a drop in $\textit{S}_{mag}$ at $H_1$,
concluding with a final entropy rise above $H_2$. The magnetoresistance also
established this three-step behavior.

Long-time pulsed ($>$\,100\,ms) magnetic fields were brought to bear for
magnetization experiments reaching 44\,T at 0.5\,K \cite{Harrison03}.
These measurements revealed that the HO phase is destroyed before
the appearance of the reentrant phase between 36 and 40\,T.
At temperatures above the maximum in the ``novel'' phase, $\approx 6$\,K,
the magnetization and susceptibility ($dM/dH$) indicate an itinerant
electron metamagnetism, i.e., an ever-sharpening peak in the
susceptibility which disappears at the reentrant transition. Yet extrapolation
of its $T$-dependence through the novel phase suggests a quantum-critical
end point at 38\,T. Thus a putative quantum critical point is overpowered
by the formation of the novel phase.

\begin{figure}
\includegraphics[width=0.99\linewidth]{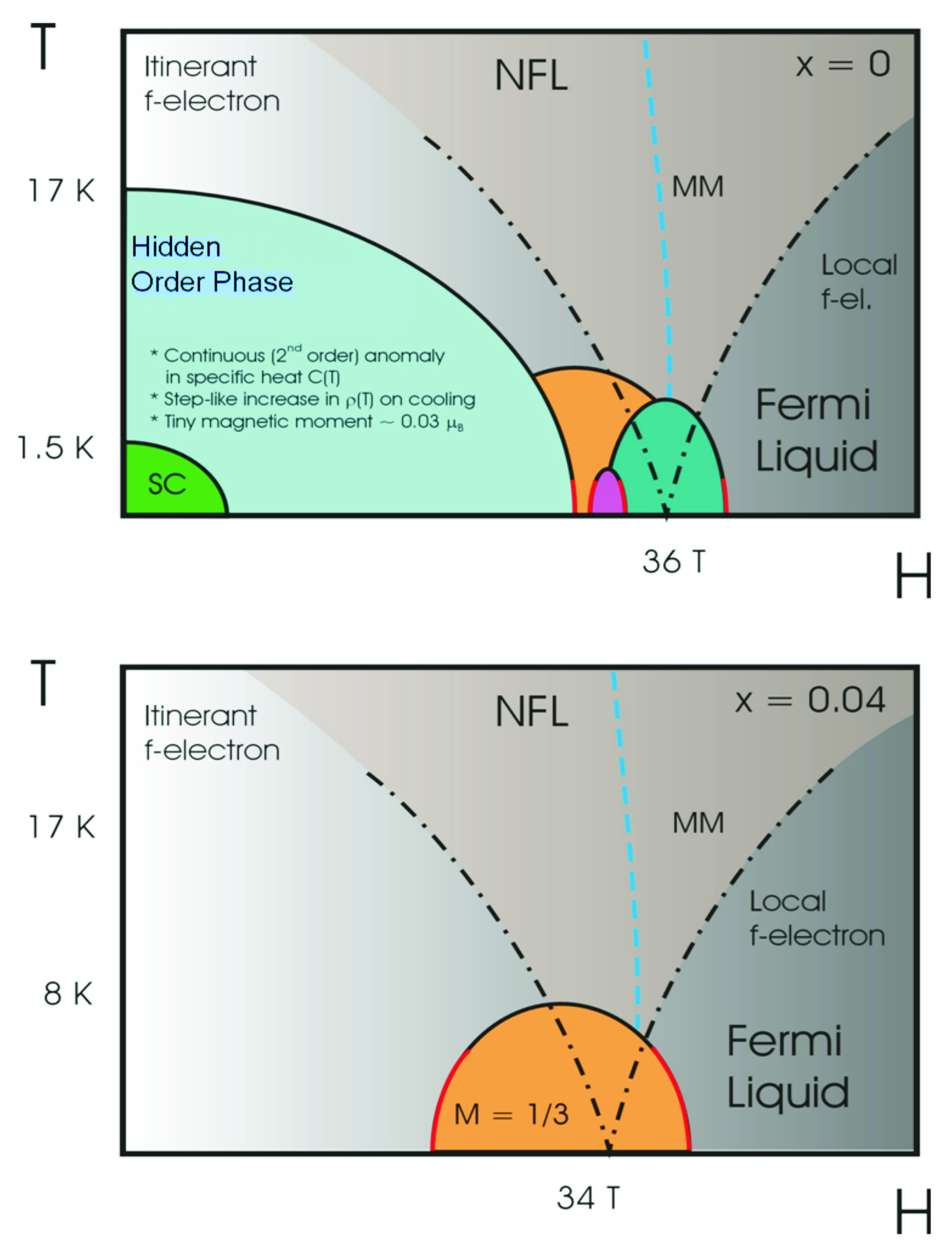}
\caption{(Color online) Sketch of the high magnetic field {$T -- H$} phase diagrams for URu$_2$Si$_2$ and U(Ru$_{0.96}$Rh$_{0.04}$)$_2$Si$_2$. MM indicates the metamagnetic transitions, {NFL the non-Fermi liquid}, and phase II is colored in orange. After \textcite{Jaime07}.} 
\label{magnetic-phase}
\end{figure}

Further experiments to examine this magnetic-field-induced critical
point were carried out by \textcite{Kimk03} via a comprehensive
magneto-resistivity study. By plotting continuous variations of $\rho (H)$ at various
temperatures and $\rho (T)$ at various fields, the extremities in $d\rho /dH$
and $d \rho /dT$ were determined; and by combining these results with the above
findings, a complete $T - H$ phase diagram could be constructed.
Figure \ref{magnetic-phase} exhibits
this high-field phase diagram with the details of the five distinct phases
three of which are the newly discovered novel phases.

Recent measurements of quantum oscillations in the resistivity, {i.e., 
the} Shubnikov-de Haas {(SdH)} effect, by \textcite{Jo07} demonstrated the
dramatic reconstruction of the Fermi surface when leaving the HO
phase and entering phase II (orange in Fig.\ \ref{magnetic-phase}) spanning the
fields region around 35\,T. The SdH results suggest an increase in
the effective carrier concentration thereby destabilizing the
gapping of the itinerant quasiparticles which leads to the
increase in the magnetization. The high-field sweeps were also
performed under pressure beginning in the LMAF phase and
field-driven to novel phase formation. Here the high-field phases
determined by resistivity show qualitatively the same number and
$T - H$ shapes as the ambient pressure phases. This means that
pressurizing the material from HO to LMAF does not alter the
resulting high-field phase formations, again indicating the
similar FSs in HO and LMAF. The high-field Fermi surface
reconstruction was further indicated from a combination of
resistivity, Hall and Nernst effect measurements
\cite{Levallois09}.

The effect of Rh-doping allows us to simplify the high-field phase diagram. It has been known for since the early work of \textcite{Amitsuka88}
that small amounts of Rh substituted for Ru create puddles of LMAF and suppress the HO via a combination of stress at the larger Rh-sites and the
addition of an extra $4d$ electron. Above 4 percent Rh substitution on the Ru site both HO and LMAF are removed; and there is only a heavy Fermi
liquid (HFL) ground state in low fields, i.e., no ordered phase of any kind. A large magnetic field has been applied to study its effect on the
HFL. \textcite{Kim04} carried out a systematic study of Rh substitution in U(Ru$_{1-x}$Rh$_x$)$_2$Si$_2$ in fields up to 45\,T. The complicated five
phases were reduced at $x = 0.04$ to a single field-induced phase: The surviving symmetric dome-like shape of phase II (orange in Fig.\ \ref{magnetic-phase}) spans
the field 26 to 37\,T with a maximum peak temperature of 9\,K. By extrapolating the HFL-behavior of the resistivity and magnetization data from
outside the dome to inside, a single point is reached at 0\,K and 34\,T, again suggesting a field-induced quantum critical point that is suppressed
by the formation of a field-induced novel phase.
Further analysis of the 4 percent Rh doped samples, U(Ru$_{0.96}$Rh$_{0.04}$)$_2$Si$_2$,
was carried out by \textcite{Silhanek05,Silhanek06a} using specific heat and magnetocaloric measurements. Here the high-field novel phase II was
mapped out and comparisons drawn with the valency transition in (Yb$_{1-x}$Y$_x$)InCu$_4$ and the metamagnetic transformation in CeRu$_2$Si$_2$.
Figure \ref{magnetic-phase} also compares the $H - T$ phase diagram for U(Ru$_{0.96}$Rh$_{0.04}$)$_2$Si$_2$ with that of undoped URu$_2$Si$_2$. Recently a
theoretical description of the suppression of HO by Rh impurities was presented by \textcite{Pezzoli11}. They studied the local competition of
HO-$\Psi$ with LMAF-$m$ where disorder is the driving force of the two competing effects. Accordingly, the phase diagram as a function of x-Rh is
obtained and compared with the local LMAF ``patch'' model
derived from $^{29}$Si NMR \cite{Baek10}.

Another way of slowly quenching the HO with doping is to substituted Re for Ru.
A recent study \cite{Butch09} has found that $\approx 15$ percent Re substitution leads to a
ferromagnetic phase beginning at a putative quantum-phase transition. This work
suggests the possible formation of a ferromagnetic quantum-critical point.

A final aspect of doping the pure URu$_2$Si$_2$ compound is to
reduce the U concentration to the dilute limit, i.e.,
R$_{1-x}$U$_x$Ru$_2$Si$_2$ where R = Th, La; Y and $x$ is a few
percent U. Such investigations were pioneered by
\textcite{Amitsuka94} and spanned a decade of experimentation
\cite{Yokoyama02}. The important questions posed here are: Can
dilute U in a non-magnetic matrix support a single-impurity Kondo
effect or even a multichannel Kondo behavior \cite{Coxd87}?
Are the $5f$ electrons of the single-ion U localized and magnetic
with a Kondo effect or itinerant - dissolved into the
conduction density of states at $E_F$? {As is well-known in Kondo physics}
it would seem that this depends on the particular matrix, Th, La, or V  \cite{Marumoto96}.
The conclusion after many measurements was {however} ambiguous, neither of
the above two possibilities {-- conventional or multichannel Kondo effect --} 
were clearly established. Presently
this topic has gained renewed interest from both the theoretical
and experimental sector \cite{Toth10}.

\section{Present State of HO}
\label{sec:present}

There persists particularly strong interest today in solving the
HO enigma. At present (January 2011) experiment has turned
towards repeating some of the previous (20 year old) measurements
using the latest advances in experimental methods and crystal
growth as, for example, with the aforementioned INS \cite{Bourdarot10}. Here we
consider first the recent progress made in optical
conductivity and quantum oscillations. Further experimentation has
taken advantage of techniques developed for and effectively
utilized in the high temperature superconductors, {\it viz.}, STM/STS
and ARPES which we  {discuss} below. 
A surge in new theories dedicated to the HO is
simultaneously taking place; their current status is surveyed
below.

Optical conductivity-redux is currently underway on the new
generation of URu$_2$Si$_2$ crystals by two groups
\cite{Levallois10,Lobo10}. The goal here is to perform
systematic spectroscopy measurements on oriented samples down to
lower frequencies and temperatures. Such experiments have probed
the gradual formation of the hybridization gap which was
found to start {above 40 K} in the KL {(HFL)} phase reaching {a dip width of}
$\approx 15$ meV as the temperature is reduced towards $T_{\rm o}$. Also an
anisotropic reduction of the conductivity was detected between the
$a$ and $c$ axes due to the hybridization \cite{Levallois10}).
Hence, a reconstruction of the electronic structure already
{\it above} $T_{\rm o}$ is suggested. The opening of the HO gap
(expected to be $\approx 5$ meV {or 40$^{-1}$ cm}) below $T_{\rm o}$
{has still not been finalized.} Although clearly  {indicated} by \textcite{Lobo10},
in accordance with older measurements
\cite{Bonn88,Thieme95,Degiorgi97}, it {had only recently been} observed 
{down to 5 K and 1.2 meV (10 cm$^{-1}$) in reflectivity \cite{vdMarel11}.} Further experiments
are now in progress to  {systematically study the optical conductivity below the HO transition.}

\begin{figure}
\includegraphics[width=0.9\linewidth]{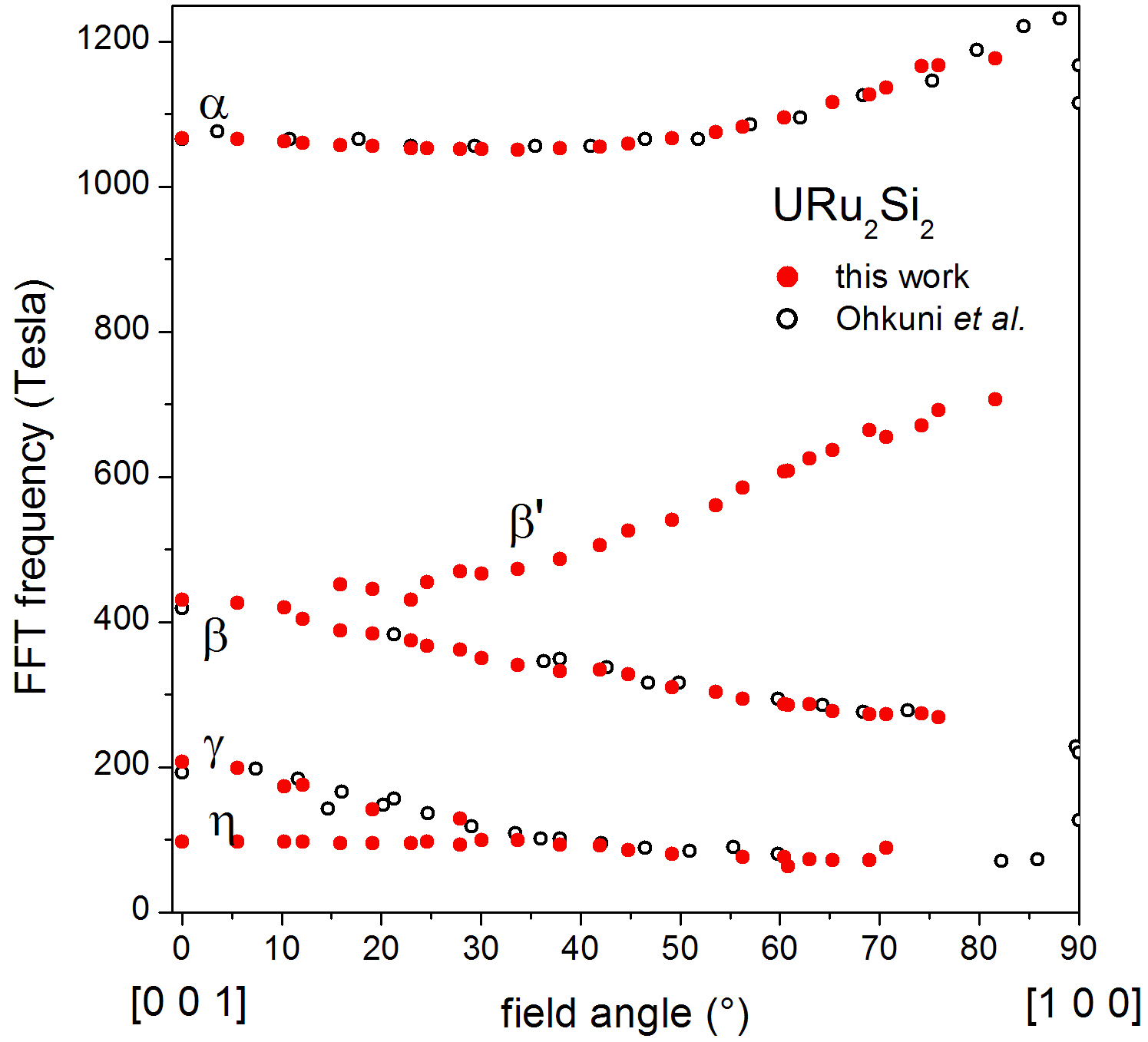}
\caption{(Color online) Angular dependence of extremal quantum oscillation {frequencies} in {\urusi}, comparing the de Haas-van Alphen data of \textcite{Ohkuni99} and the Shubnikov-de Haas data  of \textcite{Hassinger10} (labeled `this work'). A previously unseen splitting of the $\beta$ branch was observed, as well as an $\eta$ orbit that coincides almost with the previously detected $\gamma$ orbit.
Note that the large extremal orbit $\varepsilon$ was not detected (cf.\ Fig.\ \ref{fig:FS}). {FFT: fast Fourier transform.}
From \textcite{Hassinger10}.
\label{fig:SdH-Hassinger}}
\end{figure}

\begin{figure}
 \includegraphics[width=0.8\linewidth]{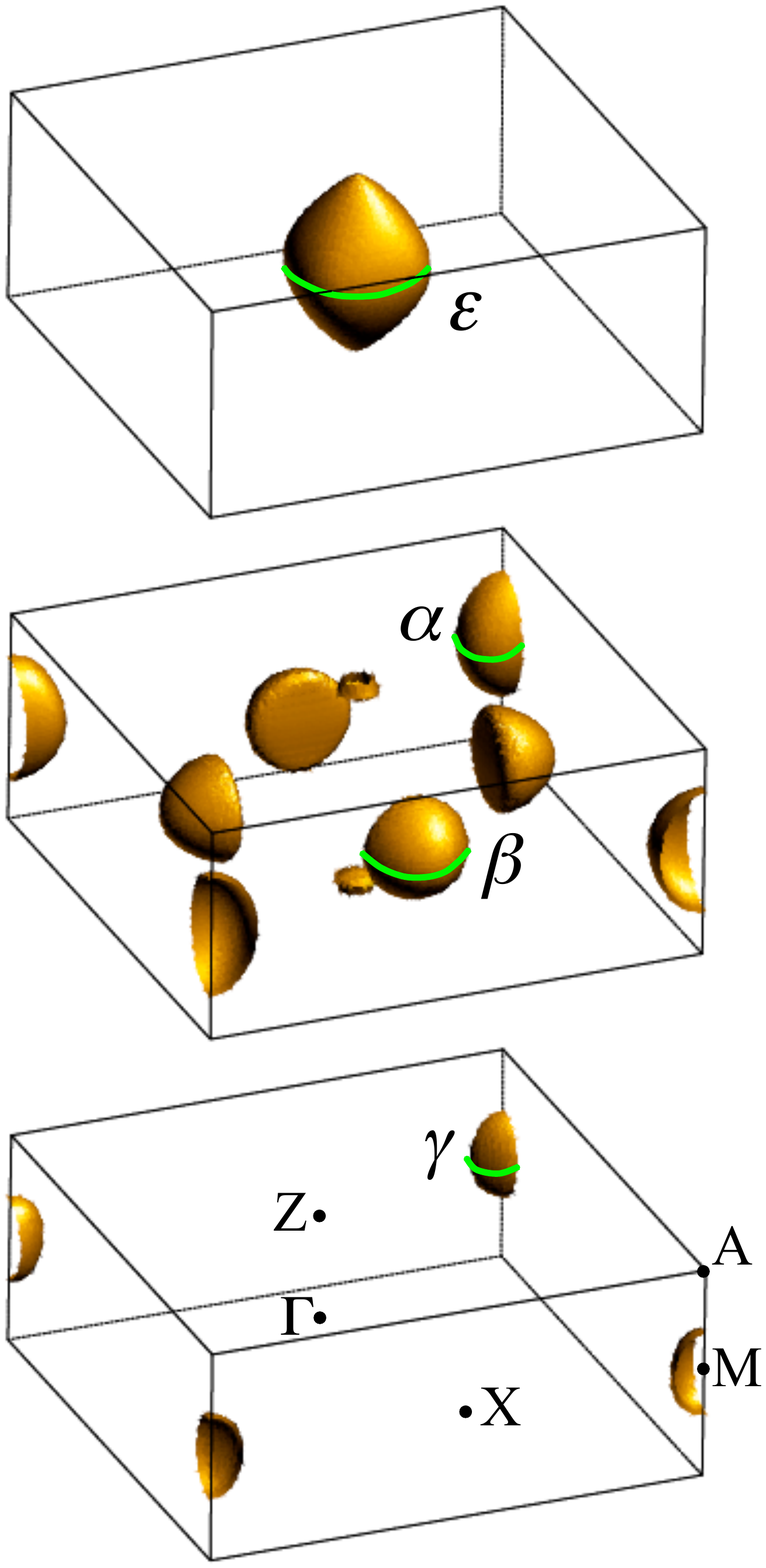}
  \caption{(Color online)  Computed Fermi surface sheets of {\urusi} in the  LMAF phase. The extremal Fermi surface orbits for field along the $z=c^{\star}$ axis are indicated by green lines,
 the branches are labeled by greek letters. Not visible is the small $\Gamma$-centered ellipsoid (called $\eta$ in Fig.\ \ref{fig:SdH-Hassinger}) that is inside the large $\Gamma$-centered surface in the top panel.  High-symmetry points are indicated in the bottom panel. After \textcite{Elgazzar09}.
  \label{fig:FS}}
\end{figure}

A new series of Shubnikov-de Haas
experiments have just been completed on high-quality single
crystals of URu$_2$Si$_2$ under pressure \cite{Hassinger10} {and
in high magnetic fields \cite{Altarawneh11}.}  {The}
fresh {angle-dependent} data {of \textcite{Hassinger10}}, shown in Fig.\ \ref{fig:SdH-Hassinger}, {reveal} two
earlier undetected FS branches in the HO phase which has increased
the observed enhanced mass to over 50 percent of that 
{derived}
from specific heat. When combined with recent SdH
measurements \cite{Shishido09}, also on a high-purity single
crystal, where another new, heavy branch
was discovered, we now believe the FS to be
definitely established. By increasing the pressure the LMAF was
reached and the quantum oscillations were measured and compared to
those in the HO state. There was very little change in the FS
between these two phases. This is unexpected, as the HO and
LMAF phases are considered to be separated by a first order phase
transition \cite{Motoyama03}. The new SdH result supports the
conjecture that the lattice doubling and modified Brillouin zone
of the LMAF is already present in the HO state. Hence, the
suggestion arises that the HO transition breaks \textsc{bct}
translational symmetry.
There exists fine agreement of the various angular dependent
FS branches  of \textcite{Hassinger10} and those predicted
from DFT calculations \cite{Oppeneer10}; the computed FS sheets
are shown in Fig.\ \ref{fig:FS} and discussed further below.
Thus, we know now the all-important FS of HO {\urusi} and at least
partially, the occurring translational symmetry breaking.

{The recent high-field SdH investigation of \textcite{Altarawneh11} finds a sequence
of four Fermi surface changes when the field is swept to 40~T. The field-induced
modifications of the SdH oscillations are interpreted as a pocket-by-pocket magnetic
polarization of the Fermi surface with increasing field, until the HO is destroyed at $\sim 35$~T.
In addition, rotating field measurements are employed to determine the effective $g$-factor.
Consistent with earlier dHvA experiments \cite{Ohkuni99} a highly angle-dependent $g$-factor is 
obtained, emphasizing the strong Ising-like anisotropy present in {\urusi}. A first explanation of the single-ion Ising feature was proposed in terms of the anisotropic Kondo model \cite{Goremychkin02}. Alternatively, in an itinerant approach it originates from the strong spin-orbit coupling of U which is particularly effective for hybridized states in the presence of a reduced lattice symmetry. This interesting Ising property has not yet received the full attention it deserves.}

\begin{figure}
\includegraphics[width=0.99\linewidth]{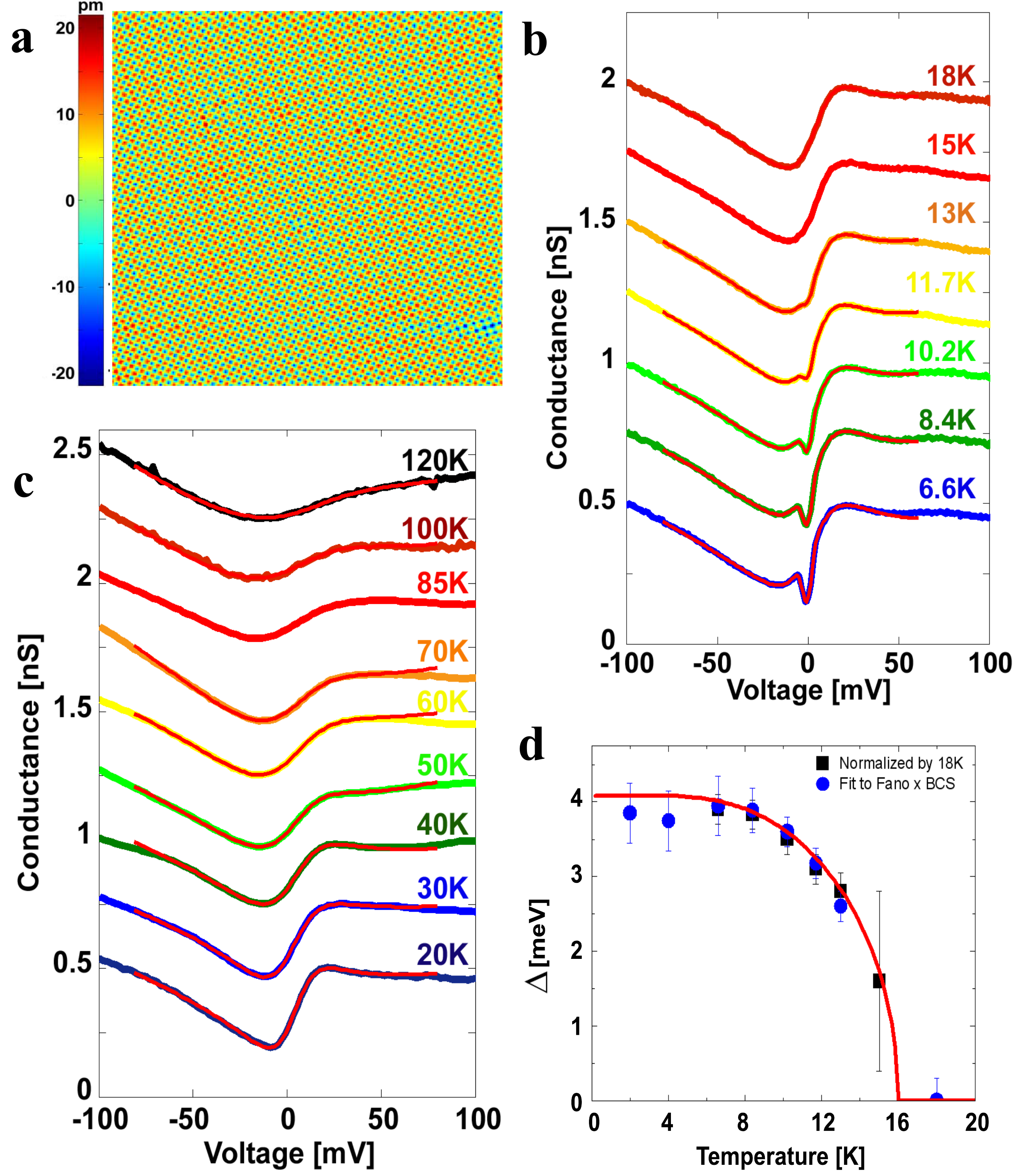}
\caption{(Color online)
STM and STS spectra of URu$_2$Si$_2$. a) Topological image of the
measured U-terminated surface (20 x 20 nm). b) Differential conductance spectrum, $dI/dV$ {\it vs}.\ $V$, spanning the HO up to $T_{\rm o}$. Note the HO gap appearance within the Fano spectrum. c) Conductance data from 120 to 20\,K illustrating the evolution of the Fano resonance.
d) Measured temperature dependence of the HO gap $\Delta$. The gap displays an asymmetric  BCS-like temperature dependence.   {Adapted from}  \textcite{Aynajian10}.}
\label{STS}
\end{figure}

Scanning tunneling microscopy and spectroscopy (STM/STS) have finally succeeded to detect the atomically-resolved properties of the heavy-fermion
material URu$_2$Si$_2$. Two groups: Cornell \cite{Schmidt10} and Princeton \cite{Aynajian10} have mastered the art of in-situ, cold-cleaving the
compound in cryogenic vacuum, thereby obtaining many tens of nanometer clean, flat surfaces for topology and spectroscopy. The issue of surface
termination remains unresolved with \textcite{Schmidt10} suggesting a Si top layer. In contrast, based upon many cleaves with different surface
reconstructions and their interface steps, \textcite{Aynajian10} have determined a U surface termination. Once the surface topology was
established, the spectroscopy could be performed with atomic resolution to detect locally modulated structures. Figure \ref{STS} shows both
the topology and the spectroscopy of the atomically resolved U surface termination.
Surprisingly, as the temperature was lowered into the Kondo
liquid regime, $T < 100$\,K, a Fano lineshape developed in the differential conductance, $dI/dV$, Fig.\ \ref{STS}c. Previously this
asymmetric line shape had been associated with a single-impurity Kondo resonance due to the two interfering tunneling paths: one through the
itinerant electrons, the other through the Kondo resonance, now the U-$5f$ electrons. The width of the Fano spectrum which varies as a function of
temperature gives an estimate of the Kondo temperature, here $\approx 120$\,K. The appearance of a Fano line shape in URu$_2$Si$_2$ has stimulated
a  {mixture} of theoretical descriptions \cite{Haule09,Maltseva09,Yang09,Figgins10,Dubi10,Wolfle10,Yuan11}.
Upon further lowering the temperature, $T <17$\,K, into the HO
state a clear dip emerges within the Fano structure, i.e., the HO
gap $\Delta_{\rm HO}$ of $\approx 5$ meV appears
\cite{Aynajian10,Schmidt10}.
Note the evolution of the Fano resonance below 100\,K and the
sharp-in-temperature appearance of the HO gap within the Fano
spectrum, Fig.\ \ref{STS}b.
$\Delta_{\rm HO} (V,T)$ is asymmetric with respect to the
Fermi energy and, most importantly, its
temperature dependence mimics a BCS mean-field gap opening.
At the lowest temperatures, $T<4$\,K (see Fig.\ \ref{STS}b), additional
structure or modes appear within the HO gap presenting a new
demand for the theory (cf.\ \textcite{Dubi10}). By taking
advantage of the atomic resolution the spatial modulations in the
conductivity can be studied. Both the Fano spectrum and the HO gap
are strongest between the U surface atoms. Thus it seems that the
main tunneling processes or largest tunneling density of states
are into hybridized electronic states that exist in-between the
U-sites, i.e., involving non-localized or itinerant $5f$
electrons.

By placing Th impurities on the U-surface, \textcite{Schmidt10}
 {were} able to image quasiparticle interference patterns. When
Fourier transformed, the bias voltage {\it vs.} the
$\textbf{k}$-space structure showed the splitting of a light band
into two heavy bands with temperature falling below $T_{\rm o}$.
Once again the splitting is of order of 5 meV at a ${\bf Q}$ = 0.3\,$a^{\star}$,
consistent with the ensemble of other measurements.

The surge of efforts in the angular resolved photoemission
spectroscopy (ARPES) field is truly remarkable since eight
international groups are presently involved in such challenging
measurements on URu$_2$Si$_2$. Although ARPES is a most successful
tool for studying the high temperature superconductors, it has
been only moderately effective for the strongly correlated
electron systems because of their 3D Brillouin zones, their
hostile cleaving and surface properties, and the need {for} very low
temperatures, ultra-high vacuum and extreme meV resolution. Yet
early on two ARPES investigations \cite{Ito99,Denlinger00,Denlinger01}
in the Kondo liquid (paramagnetic)
phase attempted to compare the the energy bands mapped
from ARPES spectra with those of the band-structure
calculations then available \cite{Yamagami98}. Here the
comparisons were of poor success without a direct determination of
or correspondence to the U-$5f$ bands near and crossing the FS.

A recent ARPES study by \textcite{Santander09} using He-lamp
energies ($\approx 21$ eV) and a high-resolution analyzer has
detected and tracked a narrow band of heavy
quasi-particles that shifts from above to below the Fermi
level as the temperature is reduced through the HO
transition. Above $E_F$ the narrow band appears
incoherent, yet once below the Fermi level the band sharpens and
disperses as a heavy-electron band that seems to be hybridized
with a light-hole band at specific $\textbf{k}$-values
corresponding to the FS of this conduction-electron band. The
measured behavior is interpreted to represent a new type
of Fermi-surface instability associated with the reconstruction of
portions of the FS involving heavy quasi-particles. This
work suggests that KL coherence develops at the HO transition and
thus not at the coherence temperature $T^{\star} \approx 70$\,K.
The specific data treatment applied might however play a role
here. Hence,  this pioneering experiment { has given rise to}
 an array of
critical questions concerning the data collection method and
analysis. Nonetheless, \textcite{Santander09} have demonstrated
the effectiveness The of ARPES in studying the HO transition in
URu$_2$Si$_2$ and their work has led to the present wave of
additional ARPES experiments.

Recently, \textcite{Yoshida10} performed laser ARPES using a 6 eV laser source on {\urusi} and U(Ru$_{1-x}$Rh$_x$)$_2$Si$_2$, with $x=0.03$.
The 3$\%$ Rh substitution is sufficient to eliminate the HO. This ARPES study,
 which is mostly sensitive to the $d$-bands, reveals that a narrow, yet dispersive band suddenly {\it appears}
below $E_F$ when temperature is lowered to below $T_{\rm o}$, a feature which is absent in the Rh-substituted sample. The sudden appearance of
this band provides evidence for a doubling of the unit cell along the $c$ axis in the HO,  an observation that agrees with that deduced  from the
latest SdH measurements \cite{Hassinger10}. 

ARPES in the soft-X-ray range was also recently reported
\cite{Kawasaki11a,Kawasaki11b}. In this energy range photoemission probes the bulk electronic structure, whereas it is particularly surface sensitive at He~I and He~II energies.
Variation of the photonenergy reveals $5f$-related energy bands dispersing in an energy window 
from $-0.6$ eV to $E_F$, which are in good agreement with band-structure calculations assuming itinerant $5f$ electrons. Moreover, these bulk sensitive ARPES measurements do not detect a narrow heavy band in the vicinity of $E_F$, in contrast to the He~I study of \textcite{Santander09}. This narrow heavy band is consequently attributed to a surface contribution.

{ 
A new technique, time and angular resolved photoemission spectroscopy
(tr-ARPES) was very recently applied to study the dynamics of the HO quasiparticles on the femtosecond time scale \cite{Dakovski11}. Within reciprocal space the probed position is  first tuned to one of the Fermi surface hot spots (see below), and subsequently the femtosecond time-resolved quasiparticle dynamics at this location is probed after stimulation with  a pump laser. Measurements performed below and above $T_{\rm o}$ reveal that the quasiparticle lifetime dramatically rises with one order of magnitude upon entering the HO. The formation of long-lived quasiparticles at the hot spots is identified as the principal mediator of the HO phase. Although the study proved the existence of driving quasiparticles, their
precise nature could not yet be disclosed.
}

\begin{figure}[tbh]
   \includegraphics[angle=-90,width=0.9\linewidth]{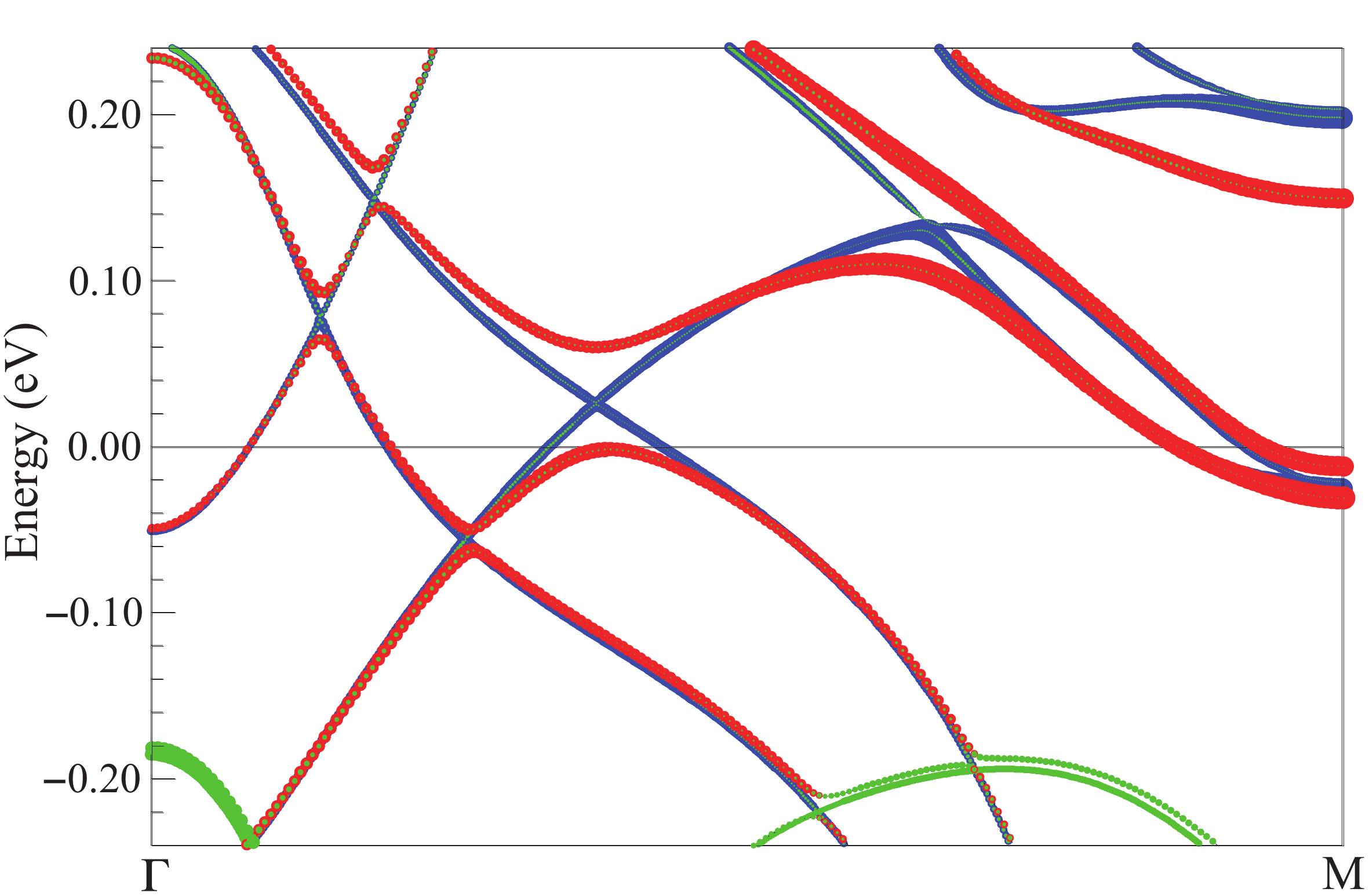}
   \includegraphics[width=0.7\linewidth]{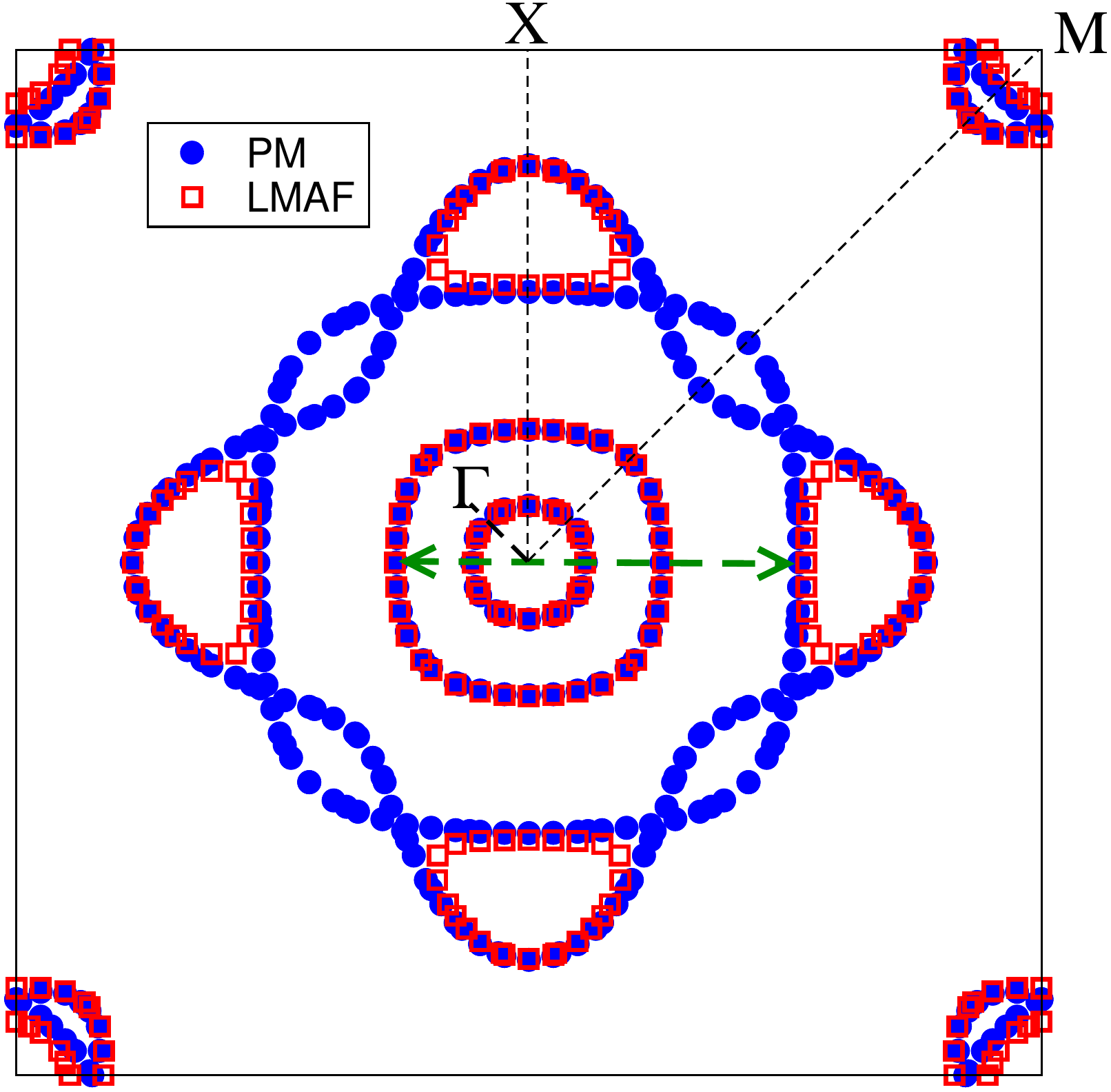}
  \caption{(Color online) Top: computed energy dispersions of {\urusi} along the $\Gamma -$M high-symmetry direction in the \textsc{st} Brillouin zone. The character of the states responsible for the bands is shown through the colors, where blue color illustrates U $5f$ character in the paramagnetic (PM) {normal state}  and red color U $5f$ character in the LMAF phase. Green color depicts the Ru $4d$ character. The amount of Ru $d$ or U $f$ character is provided through the thickness of the  bands. Note that energy dispersions for the PM {normal state} and LMAF phases are almost on top of each other, but a conspicuous lifting of a band degeneracy and a concomitant opening of a gap occurs between $\Gamma$ and M.
Bottom: cross sections of the PM (blue circles) and the LMAF (red squares) phase in the $k_z =0$ plane, showing the removal of FS portions in the LMAF phase: the blue FS sections between the  rounded half-spheres becomes completely gapped. Other FS are parts are not affected, the LMAF and PM FS cross sections are almost identical. The dashed arrow indicates the nesting vector $0.4\,a^{\star}$ (i.e., $(1\pm 0.4,\,0,\,0)$ in the \textsc{bct} structure).
  {Adapted} from \textcite{Oppeneer10}.
   \label{fig:Bands}}
\end{figure}

Contemporary theory has revisited the early band-structure
calculations {\cite{Norman88,Rozing91}} 
with the use of state-of-the-art electronic structure
methods. \textcite{Elgazzar09} performed such calculations on
URu$_2$Si$_2$ within the framework of density functional theory
(DFT) within the local density approximation (LDA). While such an
approach may be suspect for a strongly correlated electron system,
it represents the basic starting point for more sophisticated
methods. The calculation was applied to the paramagnetic (or Kondo
liquid) phase and the LMAF phase, in which there is a
reasonable magnetic moment ($\sim 0.4$ $\mu_B$) and a conventional
antiferromagnetic transition. Here the computed band structure and
FS obtained from the energy dispersions showed critical regions or
``hot spots" in {\textbf{k}}-space where degenerate band
crossings at $E_F$ (``Dirac points") induce FS instabilities.
These accidental band degeneracies {in the normal state}, shown in Fig.\
\ref{fig:Bands} {by the blue lines}, causes FS gapping when symmetry breaking
through {antiferro}magnetic ordering takes place {(red lines)}. Thus we have the LMAF
phase transition. An increase in the exchange interaction leads to
a larger magnetic moment and a greater FS gap.

Applying various computational methods to treat the $f$
electron correlations, {\it viz.}\ DFT-LDA, LDA with added strong
Coulomb interaction (LDA+$U$), and dynamical mean field theory
(DMFT), an in-depth comparison of calculated and known
experimental properties of the PM and LMAF phases was performed by
\textcite{Oppeneer10}. Good agreement is found throughout
when the $5f$ electrons are treated as itinerant, especially
regarding the 
ordered U-moment which is composed of opposite spin and orbital components in accord with the net measured neutron moment of 0.4 $\mu_B$
\cite{Bourdarot03b,Amitsuka07,Butch10}. Hence, the LMAF phase and its FS gapping is understood. Now there remains the difficulty with the HO phase
since this phase is not long-range magnetically ordered, i.e., there is no intrinsic HO magnetic moment.
 Yet we know from experiment that the
HO and LMAF phases have quite similar properties, for
example, the transport energy gap \cite{Jeffries07} and the FS
\cite{Nakashima03,Hassinger10}.
Therefore, the band-structure calculations should also apply to the HO state, provided appropriate assumptions are made for the origin of the HO.
In this phase there are the long-lived, longitudinal, dynamical spin fluctuations observed in the INS resonance. Based upon this commensurate
${\bf{Q}}_0$ spin resonance, \textcite{Elgazzar09}
proposed a novel dynamical symmetry breaking model. They argued that the antiferromagnetic spin mode causes the FS gapping and thus drives the HO
transition. 
Here unit-cell doubling along the $c$ axis and time-reversal symmetry breaking due to a dynamic mode were
 proposed for the HO in the electronic structure calculations of \textcite{Elgazzar09},
 leading to the FS topology shown in Fig.\ \ref{fig:FS}.
 Detailed DFT calculations of quantum oscillations have recently been performed
 \cite{Oppeneer10}. The calculated extremal FS
cross-sections of the theoretical FS of Fig.\ \ref{fig:FS} are in good
agreement with recent experimental data. There are five branches; four of these
correspond to the branches detected by \textcite{Hassinger10}, including the
splitting of the $\beta$ branch. Theory also predicts a larger $\varepsilon$ branch
(1.35 kT) that was detected by \textcite{Shishido09}, but
not by \textcite{Hassinger10}.
%
The commensurate spin mode drives the FS gapping and leads to the average gap magnitude as
the order parameter and the integrated intensity of the resonance
 (dynamical susceptibility) as a secondary order parameter.
Both order parameters  exhibit a BCS-like temperature
dependence. A new aspect now enters the theory, namely, the
time-scale of the spin resonance \cite{Oppeneer10}. However, the
central question persists: Can a dynamical mode create a phase
transition?

\textcite{Balatsky09} proposed that the incommensurate spin mode at ${\bf Q}_1$ would be essential to the HO instead of the antiferromagnetic mode
at ${\bf Q}_0$. Assuming a coupling between FS sheets connected by the nesting vector ${\bf Q}_1$, giving rise to a FS gap in the HO, they derived
a form of the dynamic spin susceptibility $\chi ({\bf Q}_1, \omega )$ in accordance with INS measurements. Also, \textcite{Balatsky09} calculated
the temperature-dependent specific heat related to the assumed FS gapping and obtained good agreement with experiment. INS studies of the
commensurate  spin resonance were reported recently \cite{Bourdarot10}, similar studies at the incommensurate mode are now required to unveil
their distinct contribution and relative importance.

\textcite{Haule09} have presented a new theoretical approach to
calculate the correlated electronic structure of URu$_2$Si$_2$ and
related it to the known experimental facts. Using a
combination of DFT and temperature dependent DMFT applied to a
U$^{4+}$ - $5f^2$ localized configuration, the authors deduce a
CEF scheme consisting of the $\Gamma_1^{(1)}$ and $\Gamma_2$
singlets, with a CEF splitting of $\approx 35$ K. This level
scheme supports a complex order parameter, consisting of a dipolar
order $J_z$ and {simultaneously} a hexadecapolar order, $(J_xJ_y + J_y J_x)(J_x^2
-J_y^2)$. The two-singlets scheme and possible multipoles are
those that were considered earlier by \textcite{Nieuwenhuys87} and
\textcite{Santini94}, except for the reversed order of ground and
first excited state. Upon lowering the temperature the system
evolves from this local $5f^2$ configuration into a multichannel
Kondo state which at temperatures less that 35\,K becomes
``arrested'' by the CEF splitting. The two phases, HO and LMAF,
are treated as real, respectively, imaginary, parts of the same
OP, and arise from collective CEF excitations to the excited
state. Accordingly, the HO phase transition is the real part
formation of the non-magnetic hexadecapole OP that breaks
rotational symmetry but preserves time-reversal symmetry. Upon
tuning, e.g., pressure, the imaginary part of the OP dominates and
corresponds to the LMAF phase with its time-reversal symmetry
breaking. Thus this complex OP treats both phases with their
different symmetry breaking properties.
%
 \textcite{Haule09} also calculate the one-electron
spectral function which they correlate with the recent STM/STS
experiments \cite{Aynajian10,Schmidt10}; and their results give
the observed Fano resonances. The microscopic DMFT calculations
 furthermore determine the FS and DOS. These
quantities are however disparate from those obtained in another
recent DMFT calculation that started from a nearly itinerant $5f$
configuration plus a modest Coulomb interaction $U$
\cite{Oppeneer10}. The FS predicted for a $5f^2$ hexadecapolar
state also does not seem to be in correspondence with the recent SdH results
\cite{Hassinger10}.

In order to draw further experimental comparisons \textcite{Haule10} used the above framework to develop a Landau-Ginzburg theory of the HO in
URu$_2$Si$_2$. Here they consider phase modifications due to strain, pressure and magnetic field as has been studied experimentally via uniaxial
stress \cite{Yokoyama05}, thermal expansion \cite{Motoyama08} and neutron scattering \cite{Aoki09}. The theory also associates the INS
${\bf{Q}}_0$ resonance with a pseudo-Goldstone mode which carries a fluctuating magnetic moment in the HO phase as seen in the neutron scattering.
The impact of the Haule and Kotliar theory necessitates the experimental search for some evidence of the CEF levels and their splitting, which up
until now has not been found. In addition with newly available
x-ray techniques,
e.g., x-ray Bragg diffraction \cite{Lovesey05} or
non-resonance inelastic x-ray scattering, the observation of
high-order multipolar transitions might be a future possibility
\cite{Gupta10}.

A further and different approach by \textcite{Harima10} was to present a space group analysis of the URu$_2$Si$_2$ crystal structure and thereby
approach the HO transition as a change of crystal symmetry. Based upon group theoretical tabulations they find that a second-order structural
transition occurs from the ``mother'' space group $I4/mmm$ to $P4_2 /mnm$ which does not require a lattice distortion and would keep the NQR
frequency at the Ru site unchanged \cite{Saitoh05}.
The analysis is performed within a localized, quadrupolar
framework, adopting the $\Gamma_5$ doublet to form
antiferro $J_xJ_y$ quadrupoles. In order to detect the proposed
charge distribution of quadrupole pairs, greatly improved resonant
x-ray sensitivities are needed since up until now there is no firm
experimental evidence for quadrupolar formation or its antiferro
long-range order \cite{Amitsuka10,Walker11}.

Along similar lines \textcite{Su11} have returned to predict a CDW
model using the theoretical results of the hybridization wave
calculation \cite{Dubi10}. Here the primary OP is caused by
an incommensurate hybridization between light and heavy fermion
bands near the Fermi level at ${\bf{Q}}$ = $\pm$ 0.3\,$a^{\star}$. The resulting
scattering between $f$-electrons and $d$-holes at $\pm$
${\bf{Q}}$ generates an instability forming a hybridization wave in
momentum space. Experimental comparisons are drawn with the Fano
resonance and HO gapping from STM/STS \cite{Aynajian10,Schmidt10}.
The CDW is in this scenario induced as a secondary order parameter by the
primary HO one. We note that during many years a vast experimental
scrutiny to detect a CDW has failed. So at the present moment with
many innovative efforts the search continues for the solution of
the unsolved case of HO in URu$_2$Si$_2$.

Recently, \textcite{Okazaki11} performed magnetic
torque measurements on {\urusi} to determine the magnetic
susceptibility in the tetragonal basal plane. The authors observed
a rotational symmetry reduction, from four-fold in the
paramagnetic phase to two-fold for temperatures below $T_{\rm o}$.
They interpret this as evidence of 
a non-zero off-diagonal magnetic susceptibility $\chi_{xy}$ in the HO.
The torque effect can be found only in one or two tiny single crystals of {\urusi}.
 \textcite{Okazaki11} propose that
domain formation prevents the detection for
larger crystals. According, this breaking of four-fold rotation symmetry in a tetragonal crystal is related to an electronic nematic phase, i.e.,
a directional electronic state in a heavy fermion metal (see, e.g., \textcite{Podolsky05} for a discussion of spin nematic phases). The possibility of a reduction from tetragonal symmetry was earlier investigated by in-plane oriented thermal expansion measurements, but these could
not detect any notable effect, $\Delta L /L < 10^{-7}$ \cite{Kuwahara97}. 

\textcite {Thalmeier11} have developed a Landau free energy functional
to describe the possible variety of multipolar order parameters that could be compatible with the torque measurement. Their conclusion is that an E-type, (antiferro-) two-component
 ($O_{yz},O_{zx}$) quadrupole
can best fit the two-fold torque oscillations as the HO symmetry.
In addition, based upon the torque results \cite{Okazaki11},
\textcite{Pepin-unp} have proposed spatially modulated spin-liquid
order in the basal plane as the HO. Such order is created by a Kondo
breakdown critical point.
{ 
\textcite{Fujimoto11} examines scenarios for a spin nematic state in an effective two-band model with nesting properties
of the two bands as given by first-principles calculations. The spin nematic phase is proposed to be a spin-triplet electron-hole pairing state, with electron and hole particles, respectively, located in either one of the two nested bands. 
This leads to a $d$-type pairing OP that does not break time-reversal symmetry but has broken four-fold symmetry in the basal
plane. \textcite{Oppeneer11} proposed that the nonzero off-diagonal susceptibility is caused by dynamical spin-orbital currents circulating around U atoms in the tetragonal planes.} 

These new results  {emphasize} the question {of} which symmetry is spontaneously broken in the HO. The recent laser-ARPES experiments \cite{Yoshida10}
suggest doubling of the unit cell along the $c$ axis, in agreement with SdH data which trace a very similar FS in both the HO and LMAF
\cite{Hassinger10}.  The latest torque measurements suggest a symmetry reduction in the tetragonal plane \cite{Okazaki11}.  A major
question that persists is whether time-reversal symmetry is broken in the HO. Experiments have not been able to conclusively answer this question.
Early on, the spurious SMAF phase obscured its clarification and even current experiments have not provided the final answer.  $\mu$SR experiments
performed in the HO and LMAF phases \cite{Amato04} unveiled a peculiar difference: A strongly anisotropic dipolar field along the $c$ axis in the
LMAF phase, corresponding to Ising-like order, but a very weak, {\it isotropic} local field in the HO. This field is not related to a spurious
SMAF phase or an inhomogeneous mixing of an LMAF component, as the local field is smaller than that of 0.03 $\mu_B$ ordered moments which in
addition would be Ising anisotropic.
The presence of the internal field could point to a time-odd phase which is however apparently distinct from the long-range dipolar ordered LMAF
phase. A bothersome question is the origin of such a field. Could fluctuations of multipole states or averaging due to a spin fluctuation mode account
for this?

What have we now learned about {\urusi} and its HO phase? After a quarter century of intensive investigations seeking to uncover the HO,
the identity of this mysterious phase remains unresolved. Yet our understanding of this intriguing heavy-fermion material and the HO has markedly
increased. In 2010 finally two quantities, the FS gap and the dynamical spin susceptibility, were proven to display mean-field-like OP behavior
\cite{Bourdarot10,Aynajian10}. These findings should thus provide a hint for where to search for the OP. Quantum oscillation measurements
established a seemingly complete picture of the FS in the HO \cite{Shishido09,Hassinger10}. The surprisingly large entropy release at the HO
transition could be attributed to a gapping of intensive  {long-lived} spin excitations \cite{Wiebe07} that appear to play a critical role
\cite{Oppeneer10}. Here there clearly emerges the dominance of non-local, itinerant $5f$ electrons. In addition, at least part of the spontaneous
symmetry breaking in the HO has been discovered, as there are now clear indications for unit-cell doubling along the $c$ axis \cite{Yoshida10} and
a first observation of four-fold rotational symmetry in the basal plane \cite{Okazaki11}. The high-field and doping behaviors illustrate the
fragility of the HO and its uniqueness among heavy-fermion materials. Some of these new observations do call for further investigation and
independent confirmation, and moreover, it needs to be clarified how the various features of the HO are interconnected. We anticipate that the HO
problem will continue to raise questions and challenge our understanding of how new ordered phases of matter can spontaneous emerge.


\begin{acknowledgments}
We gratefully acknowledge discussions with J.\ W.\ Allen, H.\ Amitsuka, P.\ Aynajian A.\ V.\ Balatsky, N.\ Bernhoeft, M.\ Biasini, F.\ Bourdarot,
W.\ J.\ L.\ Buyers, R.\ Caciuffo, P.\ Chandra, P.\ Coleman, N.\ J.\ Curro, J.\ D.\ Denlinger, A.\ de Visser, T.\ Durakiewicz, S.\ Elgazzar, J.\
Flouquet, M.\ Graf, H.\ Harima, N.\ Harrison, E.\ Hassinger, K.\ Haule, M.\ Jaime, V.\ Janis, J.\ R.\ Jeffries,  G.\ Knebel, G.\ Kotliar, G.\ H.\
Lander,  N.\ Magnani, M.\ B.\ Maple, D.\ van der Marel, Y.\ Matsuda, K.\ McEwen,
 Y.\ \={O}nuki, R.\ Osborn, C.\ P{\'e}pin, C.\ Pfleiderer, J.\ Rusz, A.\ Santander-Syro, J.\ Schoenes, M.-T.\ Suzuki, P.\ Thalmeier, C.\ M.\ Varma, and A.\ Yazdani. This work has been  {supported} through the Swedish Research
Council (VR).
 \end{acknowledgments}

{\it Note added in proof.} ---After submission of this article several works on {\urusi} appeared. \textcite{Malone11} measured its thermoelectric coefficients in a high magnetic field and suggest that changes of the Fermi surface topology occur deep in the HO phase at high fields. \textcite{Liu11}  employed ultrafast pump-probe optical spectroscopy to monitor the response to a fs-laser pulse. The decay of the optical pumped state suggests the opening of a pseudogap below 25 K. Theoretical support for this idea was provided by \textcite{Haraldsen11}  who deduced the presence of a pseudogap state from PCS measurements. \textcite{Toth11} proposed a localized hexadecapolar Kondo effect in diluted {\urusi}.  Further, the most recent experimental studies on {\urusi} are by \textcite{Niklowitz11}, who performed quasielastic scattering from the ${\bf Q}_0$ and ${\bf Q}_1$ resonances, \textcite{Bourdarot11}, who performed neutron scattering under uniaxial stress, and \textcite{Nagel11} who made a Fermi liquid analysis of the optical conductivity.

\bibliographystyle{apsrmp}

\newpage



\begin{figure}[ht]
\centering
\subfigure[High T]{
\includegraphics[width=0.99\linewidth]{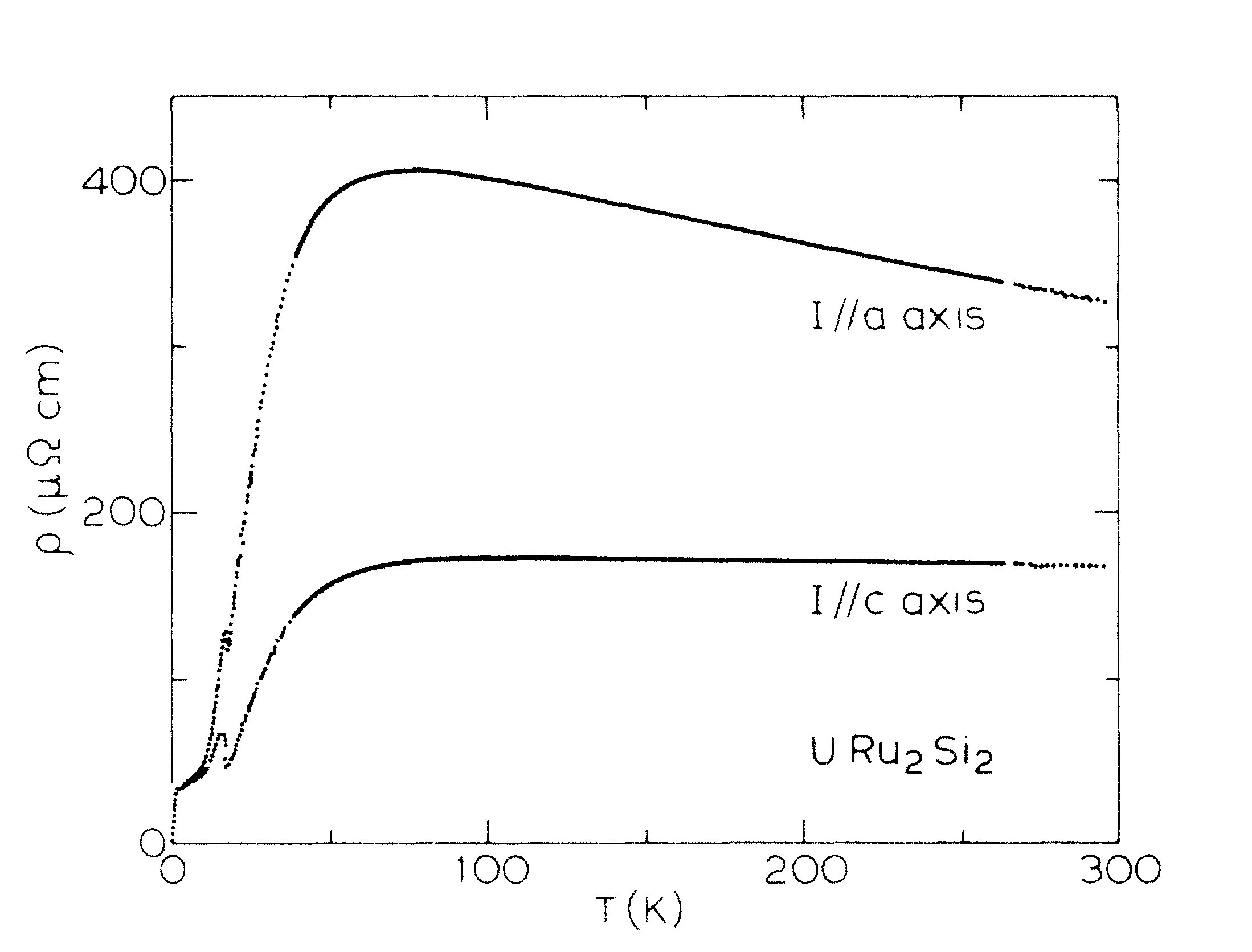}
}
\\
\subfigure[Low T]{
\includegraphics[width=0.98\linewidth]{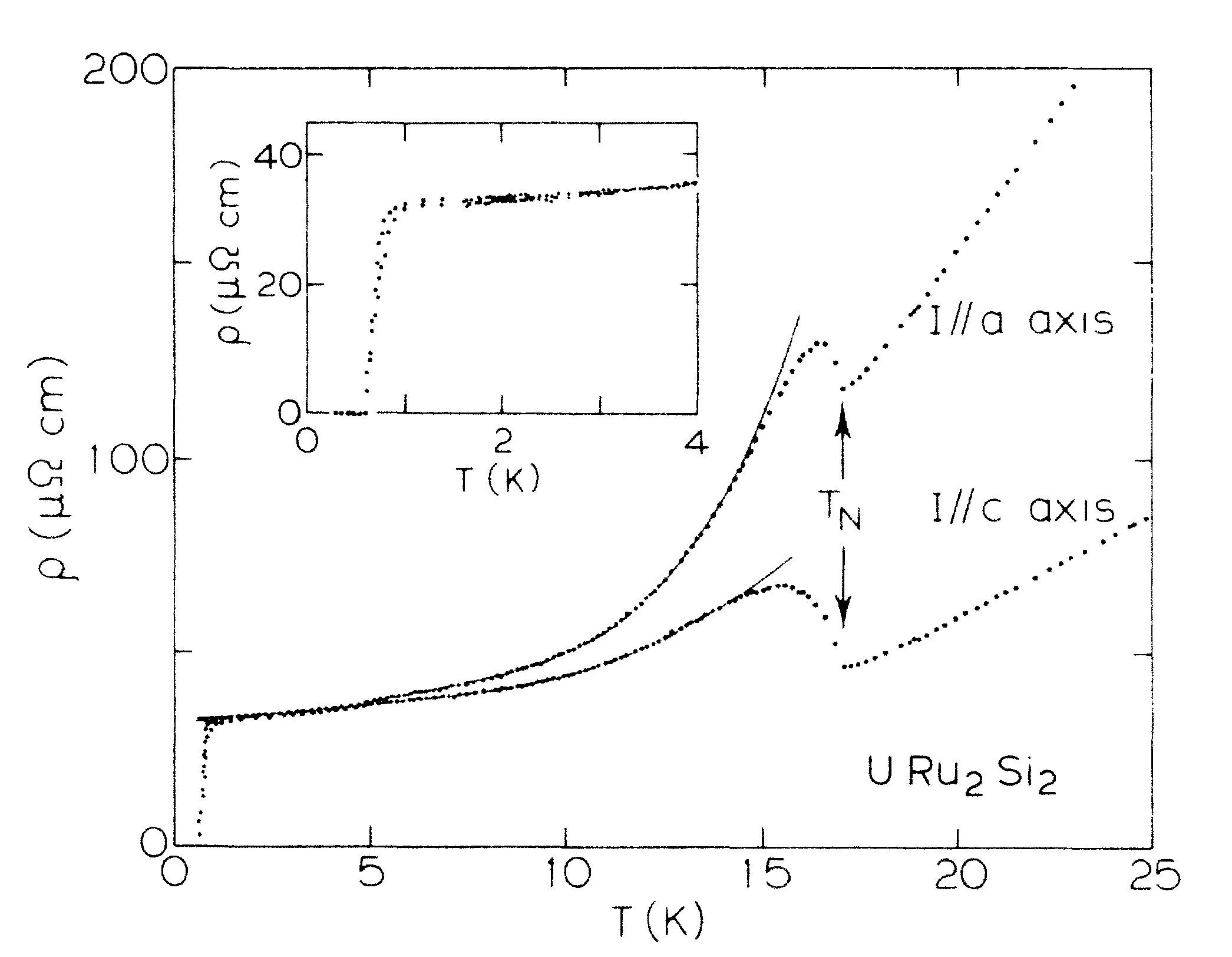}
}
\caption{Top: Overview of the resistivity {$\rho$} along the $a$ and $c$ axes.
Bottom: Expanded view of the
low-temperature resistivity illustrating the HO transition ($T_{\rm o} = 17.5$\,K) and the superconducting one ($T_c = 0.8$\,K).
After \textcite{Palstra86}.}
\label{resistivity}
\end{figure}

\end{document}